\documentstyle[epsfig,12pt]{article}
\begin{document}

\hoffset = -1truecm
\voffset = -2truecm
\baselineskip = 10 mm

\title{\bf Prediction for unintegrated parton distributions}

\author{{\bf Jianhong Ruan} and {\bf Wei Zhu}\\
\normalsize Department of Physics, East China Normal University,
Shanghai 200062, P.R. China \\
}

\date{}

\newpage

\maketitle

\vskip 3truecm

\begin{abstract}

   Unintegrated parton distributions in the proton and nucleus
are predicted by a modified DGLAP equation incorporating the
shadowing corrections, which include exact energy-momentum
conservation in each splitting and fusion vertices. We find that the
nuclear shadowing effects are obvious, although they are far from
the saturation limit. On the other hand, we point out that the
suppression of the unintegrated gluon distribution toward lower
$k_t^2$ may arise from the valence-like input rather than the
saturation effects.

\end{abstract}

\newpage
\begin{center}
\section{Introduction}
\end{center}

    Conventional parton distribution functions (PDFs) are
the basic quantities describing the inclusive cross sections of hard
processes within the QCD improved parton model. These distributions
are integrated over the transverse momentum $k_t$ of an observed
parton to the virtuality $\mu^2$ of the probe. Describing the
exclusive processes, where the transverse momentum of the produced
hadron can be observed, requires the introduction of more
complicated quantities, the so-called (two-scale) unintegrated
parton distribution functions (UPDFs), which generally depend on two
hard scales $k_t$ and $\mu$. Recently, theoretical and
phenomenological studies of UPDFs have been actively pursued since
they exactly correspond to the quantity that enters the Feynman
diagrams and relate to many exclusive or semiexclusive processes;
For a review, see Ref. [1].

    An existing evolution equation for the two-scale unintegrated gluon
distribution is the CCFM equation [2], in which the emission of
gluons during the initial cascade is only allowed in an
angular-ordered region. The corresponding quark distributions are
indirectly derived by the convolution of the gluon density with the
off-shell matrix element for boson gluon fusion. The solution of the
CCFM equation is much more complicated and has only proven to be
practical with Monte Carlo generators up to now. Moreover, the
interactions among initial partons are neglected in the derivation
of the CCFM equation. Obviously, this assumption is invalid at the
small-$x$ and low-$k_t$ regions, where the parton wave functions
begin overlap spatially. Therefore, the corrections of the initial
gluon fusion to the QCD evolution equation at small-$x$ should be
considered. One can imagine that gluon fusion suppresses or shadows
the growth of parton densities and leads the parton distribution
gradually approaching a possible limit form at $k_t<Q_s(x)$, where
the gluon fusion balances with the gluon splitting. This stable form
is called the saturation [3]. The value of $Q^2_s$ is the saturation
scale. The behavior of UPDFs near the saturation scale is important
for testing various nonlinear QCD evolution dynamics. Although some
recent works [4] used an absorptive boundary condition on the CCFM
equation to mimic the saturation effect, it is quite cumbersome to
introduce the gluon recombination kernels in the CCFM equation.

     Instead of the CCFM equation,
Kimber, Martin and Ryskin [5] proposed that the two-scale UPDFs can
be derived from the single-scale unintegrated distribution, and its
dependence on the second scale $\mu$ is introduced by using the
Sudakov factor. The single-scale unintegrated distributions can be
obtained by the following different ways: (i) the KMS scheme [6]:
the BFKL equation embodying leading log $k^2_t$ (DGLAP) corrections;
or (ii) the KMR scheme [5]: the DGLAP equation embodying leading log
$1/x$ (BFKL) corrections. Compared with the CCFM equation, the
modifications of the gluon fusion to the BFKL and DGLAP equations
are easily made.

    A modification of the gluon fusion to the BFKL equation is the BK
equation [7]. Both the BFKL and BK equations include only the
leading-order $(1/x)$ ($LLx$) contributions, and they work at the
small-$x$ range. The beyond $LLx$ modifications should be included
for the predictions in the full-$x$ range. However, it is difficult
to treat the subleading logarithmic corrections in the BFKL and BK
dynamics, except for making a roughly approximation [8].

   The correction of the gluon recombination to the DGLAP
equation was first proposed by Gribov-Levin-Ryskin and Muller-Qiu in
the GLR-MQ equation [9,10]. In this equation, the double leading
logarithmic approximation ($DLLA$) was taken, where the nonlinear
shadowing terms only kept the $1/x$ power contributions. Obviously,
the beyond $LLx$ corrections to the GLR-MQ kernel are necessary. We
have generalized the GLR-MQ equation in a modified DGLAP (MD-DGLAP)
equation, in which the evolution kernels were derived in the
full-$x$ range at the $LL(Q^2)$ approximation [11]. On the other
hand, the corrections of the BFKL effect due to the random
distribution of transverse momenta at the small-$x$ range are small
in the KMR scheme [5], and we neglect them in this work. Thus, we
may obtain the two-scale UPDFs based on the DGLAP dynamics including
the shadowing effect. This is one of the reasons why we try to
evolve the two-scale UPDFs by using the MD-DGLAP equation. Besides,
the linear DGLAP evolution and nonlinear recombination corrections
in the MD-DGLAP equation were derived by using the cutting rules
based on the time-ordered perturbation theory (TOPT), where the
contributions from the real and virtual diagrams to the MD-DGLAP
equation were naturally separated. Thus, we can easily calculate the
virtual contributions with the Sudakov form factor in the KMR
scheme.

    At a first step, we compute the MD-DGLAP equation with the data from
the HERA to obtain a set of integrated parton distributions in Sec.
2. The numerical solutions of the evolution equation depend
sensitively on the input parton distributions at some low scale
$\mu_0$. In this work, we use the GRV model [12], in which the
parton distributions are QCD radiatively generated from the
valence-like input distributions at an optimally determined scale
$\mu_0 < 1~GeV$. Therefore, the behavior of the parton densities at
$k_t^2> \mu^2_0$ is a purely perturbative phenomenon. This radiative
approach is especially important for the investigations concerning
asymmetrical behaviors of UPDFs at small $x$ and low $k^2_t$.
Therefore, in this work we use the GRV-type input distributions to
evolve the MD-DGLAP equation.

    In Sec. 3 we use the above-mentioned integrated parton distributions to
predict the UPDFs in the KMR scheme. We discussed the comprehensive
effects of the nonlinear terms in the evolution equation. To
emphasize the shadowing effect, we calculate the two-scale UPDFs in
the nuclear target. An interesting result is that we find that the
unintegrated gluon distribution is dropping toward low $k_t$ and it
arises from the valence-like input rather than the nonlinear
saturation effects. As we know that several UPDFs with the
saturation model have been used to explain the inclusive hadronic
production at the RHIC and further to predict the LHC physics, our
result provides an alternative model without the nonlinear
saturation mechanism. We will discuss this result in the last
section.

\newpage
\begin{center}
\section{Modified DGLAP equation}
\end{center}

    We determine a set of integrated parton distributions
with the QCD evolution equation by fitting the HERA measurements of
the proton structure function $F_2$. We know that the DGLAP equation
produces a rapid growth of gluon density at small $x$. The gluons
therefore must begin to spatially overlap and recombine. The
corrections of parton recombination to the DGLAP equation in Ref.
[11] were considered by summing up all possible twist-four cut
diagrams at the $LLA$($Q^2$). In the derivation of the equation, the
TOPT was used to pick up the contributions of the leading
recombination diagrams. As a consequence, the corrections of the
gluon recombination to the evolution of PDFs near scale $Q^2$ are
described by the following modified DGLAP equation:

$$Q^2\frac{dxv(x,Q^2)}{dQ^2}$$
$$=\frac{\alpha_s(Q^2)}{2\pi}\int_x^1\frac{dy}{y}P_{qq}(z)xv(y,
Q^2)$$
$$-\frac{\alpha_s(Q^2)}{2\pi}xv(x,Q^2)\int_0^1 dzP_{qq}(z),\eqno(1)$$
for valence quark distribution, where $z=x/y$,

$$Q^2\frac{dxs(x,Q^2)}{dQ^2}$$
$$=\frac{\alpha_s(Q^2)}{2\pi}\int_x^1\frac{dy}{y}P_{qq}(z)xs(y,
Q^2)$$
$$+\frac{\alpha_s(Q^2)}{2\pi}\int_x^1\frac{dy}{y}P_{qg}(z)xg(y,
Q^2)$$
$$-\frac{\alpha_s(Q^2)}{2\pi}xs(x,Q^2)\int_0^1dzP_{qq}(z)$$
$$-\frac{\alpha_s^2(Q^2)K}{Q^2}\int_x^{1/2}\frac{dy}{y}x
P_{gg\rightarrow q}(z)[ yg(y,Q^2)]^2$$
$$+\frac{\alpha_s^2(Q^2)K}{Q^2}\int_{x/2}^x\frac{dy}{y}x
P_{gg\rightarrow q}(z)[ yg(y,Q^2)]^2,\eqno(2)$$ for sea quark
distribution, and

$$Q^2\frac{dxg(x,Q^2)}{dQ^2}$$
$$=\frac{\alpha_s(Q^2)}{2\pi}\int_x^1\frac{dy}{y}P_{gq}(z)xs(y,
Q^2)$$
$$+\frac{\alpha_s(Q^2)}{2\pi}\int_x^1\frac{dy}{y}P_{gg}(z)xg(y,
Q^2)$$
$$-\frac{1}{2}2n_f\frac{\alpha_s(Q^2)}{2\pi}xg(x,Q^2)\int_0^1dzP_{qg}(z)$$
$$-\frac{1}{2}\frac{\alpha_s(Q^2)}{2\pi}xg(x,Q^2)\int_0^1 dzP_{gg}(z)$$
$$-\frac{\alpha_s^2(Q^2)K}{Q^2}\int_x^{1/2}\frac{dy}{y}x
P_{gg\rightarrow g}(z)[ yg(y,Q^2)]^2$$
$$+\frac{\alpha_s^2(Q^2)K}{Q^2}\int_{x/2}^x\frac{dy}{y}x
P_{gg\rightarrow g}(z)[ yg(y,Q^2)]^2,\eqno(3)$$ for gluon
distribution, where the unregularized DGLAP splitting kernels are

$$P_{gg}(z)=2C_2(G)\left[z(1-z)+\frac{1-z}{z}+\frac{z}{1-z}\right],\eqno(4)$$

$$P_{gq}(z)=C_2(R)\frac{1+(1-z)^2}{z},\eqno(5)$$

$$P_{qq}(z)=C_2(R)\frac{1+z^2}{1-z},\eqno(6)$$

$$P_{qg}(z)=\frac{1}{2}[z^2+(1-z)^2],\eqno(7)$$

$$C_2(G)=N=3,~~C_2(R)=\frac{N^2-1}{2N}=\frac{4}{3},\eqno(8)$$ and
the recombination functions are

$$P_{gg\rightarrow
g}(z)=\frac{9}{64}\frac{1}{x}(2-z)(72-48z+140z^2-116z^3+29z^4),\eqno(9)$$

$$P_{gg\rightarrow
q}(z)=\frac{1}{48}\frac{1}{x}z(2-z)^2(18-21z+14z^2).\eqno(10)$$ The
parameter $K$ in Eqs. (1)-{3) depends on the definition of double
parton distribution and the geometric distributions of partons
inside the target. For simplicity, we regard $K$ as a free
parameter. The factor 1/2 in the virtual terms of Eq. (3) due to the
symmetry of the Feynman diagrams is important for the cancellation
of the collinear singularities [11].

     In the derivation of the above-mentioned MD-DGLAP equation, the
TOPT cutting rules were used [11]. The contributions of the DGLAP
equation and the recombination corrections are presented in the
MD-DGLAP equation in a unified methodology. The contributions from
the real and virtual diagrams are completely separated although they
share a common evolution kernel. Thus, we can completely extract the
contributions from the virtual processes using the Sudakov from
factor in Sec. 3. This form of the linear parts in Eqs. (1)-(3) was
first derived by Collins and Qiu using the technique of cut vertices
in Ref. [13], but they can be more simply confirmed with the TOPT
cutting rules [11].

   The nonlinear parts of Eqs. (1)-(3) are the contributions from
$2\rightarrow 2$ and $1\rightarrow 3$ recombination processes. The
equation keeps the momentum conservation. Particularly, the positive
nonlinear terms contribute the antishadowing effect, while the
negative nonlinear terms lead to the shadowing correction. The
coexistence of shadowing and antishadowing in the QCD evolution
equation is a general requirement of the local momentum
conservation. The GLR-MQ equation does not consider the
contributions from the virtual diagrams of the recombination
processes. Fortunately, we have indicated that the contributions
from the virtual diagrams corresponding to $2\rightarrow 2$ and
$1\rightarrow 3$ recombination processes are canceled [11]. Thus, we
do not need to compute the virtual diagrams of the nonlinear
recombination processes in the Sudakov form factor, provided the
equation includes the contributions from both the shadowing and
antishadowing effects.

    We emphasize that the combining distribution of two gluons is
assumed to be $g^{(2)}(y,Q^2)=g^2(y,Q^2)$ either in the MD-DGLAP
equation or in the GLR-MQ equation. This is the simplest model. One
of us (W.Z) has discussed the recombination of gluons with different
values of $x$ in the nonlinear evolution equation and finds that
this modification unreasonably enhances the shadowing effect in the
GLR-MQ equation, while it does not change the predictions of the
MD-DGLAP equation, since the momentum conservation plays an
important role in this result [14].

    Since the equations (1)-(3) were derived at the leading logarithmic
($Q^2$) approximation and they contain the terms beyond the leading
logarithmic $1/x$ approximation, the modified DGLAP equation is
valid in the full-$x$ range if we neglect the BFKL corrections at
the very small-$x$ region.

    In this work we use the initial valence quark and gluon densities in
the GRV98LO set [12], that is

$$xu_v(x,\mu^2_0)=1.239x^{0.48}(1-x)^{2.72}(1-1.8\sqrt{x}+9.5x),$$

$$xd_v(x,\mu^2_0)=0.614(1-x)^{0.9}xu(x,\mu^2_0),$$ and

$$xg(x,\mu^2_0)=17.47x^{1.6}(1-x)^{3.8}.\eqno(11)$$
In the meantime, we let the parameters in the sea quark distribution
to fit the HERA data [15] using the MD-DGLAP evolution equation. As
a consequence, this input sea quark distribution in the GRV98LO,

$$xs(x,\mu^2_0)=2x[\overline{u}+\overline{d}]$$
$$=1.52x^{0.15}(1-3.6x^{0.5}+7.8x)(1-x)^{9.1},\eqno(12)$$ is
changed to

$$xs(x,\mu^2_0)=2x[\overline{u}+\overline{d}]$$
$$=0.9x^{0.01}(1-3.6x^{0.5}+7.8x)(1-x)^{8.0},\eqno(13)$$ and
the starting scale of the evolution is modified from $\mu^2_0=0.26
GeV^2$ to $\mu^2_0=0.34 GeV^2$.

    To illustrate the difference between the two input distributions, we plot in Fig. 1
our modified input sea quark distribution
$xs=2x[\overline{u}+\overline{d}]$ of Eq. (13) (solid curve) and
that of the GRV98LO of Eq. (12) (dashed curve). Our distribution is
soften than the GRV98LO set for compensating the shadowing effect in
the evolution.

    With the above-mentioned input distributions and taking $K=0.0014~GeV^2$, we evolve the
MD-DGLAP equation and take $F_2=\sum e^2_qx(s +v)$ to compare with
the HERA (H1 and ZEUS) data in Fig. 2. For comparison, we also plot
the results of the DGLAP equation with the GRV input (dashed
curves). One can see that the contributions of the gluon
recombination improve the fit at low $Q^2$.

\newpage
\begin{center}
\section{Unintegrated parton distributions}
\end{center}

    According to the definition, the two-scale UPDFs and PDFs have the following
relations:

    $$\int^{\mu^2}_0\frac{dk^2_t}{k^2_t}f_a(x,k^2_t,\mu^2)
=xa(x,\mu^2), \eqno(14)$$ where $a(x,\mu^2)=v(x,\mu^2)$,
$s(x,\mu^2)$ or $g(x,\mu^2)$. Ignoring the fact that the
unintegrated density may depend on two scales, one can roughly
estimate the unintegrated gluon distribution with

$$\frac{1}{k^2_t}f_g(x,k^2_t,\mu^2=k^2_t)\simeq\left.\frac{dxg(x,\mu^2)}{d\mu^2}\right\vert_{\mu^2=k^2_t}.
\eqno(15)$$ However, Eq. (15) cannot remain true as $x$ increases,
since the negative virtual DGLAP term may exceed the real emission
DGLAP contribution and it would give negative values for $f_g$.

    The evolutions of UPDFs on the two scales $k_t$ and $\mu$
in the KMR scheme [5] were investigated separately by the real and
virtual contributions of Eqs. (1)-(3) , thus we have

$$f_g(x,k_t^2, \mu^2)$$
$$=T_g(k_t^2,\mu^2)\left\{\frac{\alpha_s(k^2_t)}{2\pi}\int_x^{z_{max}}dz\left[P_{gg}(z)\frac{x}{z}g\left(\frac{x}{z},k^2_t\right)
+P_{gq}(z)\frac{x}{z}V\left(\frac{x}{z},k^2_t\right)+P_{gq}(z)\frac{x}{z}s\left(\frac{x}{z},k^2_t\right)\right]\right.$$
$$-\left.\frac{\alpha_s^2(k^2_t)K}{Q^2}\int_x^{1/2}\frac{dy}{y}x
P_{gg\rightarrow g}(z)\left[
\frac{x}{z}g\left(\frac{x}{z},k^2_t\right)\right]^2\right.$$
$$+\left.\frac{\alpha_s^2(k^2_t)K}{Q^2}\int_{x/2}^x\frac{dy}{y}x
P_{gg\rightarrow g}(z)\left[
\frac{x}{z}g\left(\frac{x}{z},k^2_t\right)\right]^2\right\},\eqno(16)$$

$$f_s(x,k_t^2,\mu^2)$$
$$=T_s(k_t,\mu)\left\{\frac{\alpha_s(k^2_t)}{2\pi}\int_x^{z_{max}}dz\left[P_{qg}(z)\frac{x}{z}g\left(\frac{x}{z},k^2_t\right)
+P_{qq}(z)\frac{x}{z}s\left(\frac{x}{z},k^2_t\right)\right]\right.$$
$$-\left.\frac{\alpha_s^2(k^2_t)K}{Q^2}\int_x^{1/2}\frac{dy}{y}x
P_{gg\rightarrow q}(z)\left[
\frac{x}{z}g\left(\frac{x}{z},k^2_t\right)\right]^2\right.$$
$$\left.+\frac{\alpha_s^2(k^2_t)K}{Q^2}\int_{x/2}^x\frac{dy}{y}x
P_{gg\rightarrow q}(z)\left[
\frac{x}{z}g\left(\frac{x}{z},k^2_t\right)\right]^2\right\},\eqno(17)$$

$$f_v(x,k_t^2,\mu^2)$$
$$=T_v(k_t,\mu)\frac{\alpha_s(k^2_t)}{2\pi}\int_x^{z_{max}}dzP_{qq}(z)\frac{x}{z}v\left(\frac{x}{z},k^2_t\right)
,\eqno(18)$$ where $T_a(k^2_t,\mu^2)$ is the Sudakov form factor,
resumming the virtual corrections:

$$T_g(k_t^2,\mu^2)=\exp
\left\{-\int^{\mu^2}_{k^2_t}\frac{\alpha_s(k'^2_t)}{2\pi}\frac{dk'^2_t}{k'^2_t}\left
[\int^{z_{max}}_{z_{min}}dz\frac{1}{2}P_{gg}(z)+
n_f\int^1_0dzP_{qg}(z)\right]\right\},\eqno(19)$$

$$T_s(k_t^2,\mu^2)=T_v(k_t^2,\mu^2)=\exp
\left\{-\int^{\mu^2}_{k^2_t}\frac{\alpha_s(k'^2_t)}{2\pi}\frac{dk'^2_t}{k'^2_t}\int^{z_{max}}_0dzP_{qq}(z)\right\}.\eqno(20)$$
The positive (real) terms on the right-hand sides of Eqs. (1)-(3)
describe the number of partons $\delta a$ emitted in the interval
$\mu^2< k^2_t < \mu^2+\delta \mu^2$. Such emission clearly changes
the transverse momentum $k_t$ of the evolving parton. While the
negative (virtual) contributions in Eqs. (1)-(3) do not change the
parton $k_t$ and may be resummed to give the Sudakov form factor
$T(k_t^2,\mu^2)$, the parton with transverse momentum $k_t$ remains
untouched in the evolution up to the factorization scale.

    According to Ref. [16], the strong ordering in transverse momentum
automatically ensures angular ordering and the ¡°coherence¡± effect
constraints

$$z<\frac{\mu}{\mu+k_t}\equiv z_{max},\eqno(21)$$ and

$$z_{min}=1-z_{max}. \eqno(22)$$
On the other hand, there is no ¡°coherence¡± effect for quark
emission, and therefore the phase space available for quark emission
is not restricted by the angular-ordering condition.

    We put the solution of Sec. 2 to Eqs. (16)-(18) and calculate the UPDFs.
In Figs. 3-5 we plot $x$ and $k^2_t$ dependence (solid curves) of
the two-scale PDFs in the proton at the value of the evolution scale
$\mu=10 GeV$. For comparison, we give the solutions using the DGLAP
equation with the GRV input (dashed curves). An interesting result
is that the two-scale unintegrated gluon distribution is dropping
down for $k_t^2\rightarrow 0$. We indicate the dip position of the
unintegrated gluon distribution by $k^2_D(x)$. In Fig. 6, we plot
the relation $k^2_D(x)\sim x$. We find that $k^2_D$ is small but it
still belongs to the perturbative value of $k^2_t>1~GeV^2$ for
$x<10^{-3}$, and this effect can be observed in the RHIC and LHC
energy regions. For comparison, we give the saturation scale
$Q^2_s(x)=1GeV^2(10^{-4}/x)^{0.277}$ [1] in Fig. 6.

      Two factors may suppress parton densities at low $k_t$
in the KMR scheme. Factor A is the valence-like form of the GRV
input. For example, we simplify the valence-like gluon distribution
as $xG(x,\mu^2)\sim x^{\alpha}(1-x)^{\beta}$ with $\alpha>0$, and it
implies a finite number of the initial gluons. This distribution is
drooping at $x\rightarrow 0$. On the other hand, after evolution
begins, the infinite number of the radiative gluons almost have the
distribution $xG(x,Q^2>\mu^2)\sim x^{\alpha'}(1-x)^{\beta'}$, where
$\alpha'\leq 0$. Thus, the distribution fast becomes steep at small
$x$. This immediate increase of the radiative gluons at small $x$
causes the unintegrated gluon distribution in Eq. (16) to drop down
toward small $k_t$. Factor B is the Sudakov form factor, which is
smaller than unity because of the negative contributions of the
virtual terms in the DGLAP equation. This factor is irrelative to
the parton distributions but becomes small at $k^2_t\rightarrow 0$
if $\mu^2$ is fixed.

    To confirm the above suggestion, we look at three
different examples:

    (i) The drop occurs only from factor A. For this case, we use

    $$\frac{1}{k^2_t}f_g(x,k^2_t,\mu^2=k^2_t)\equiv F(x,k^2_t),
\eqno(23)$$ to compute the single-scale unintegrated gluon
distribution. Note that it is different from Eq. (15), because the
contributions of the virtual terms have been removed. The results
are plotted in Fig. 7. Comparing Fig. 7 with Fig. 3, we find that
the decrease of $F(x,k^2_t)$ delays for $k_t\rightarrow 0$, because
the Sudakov form factor $T_g=1$ in Eq. (23).

    (ii) The drop occurs from both factors A and B. We see the $k_t$ dependence of the
two-scale unintegrated sea quark distribution using the DGLAP
equation and the GRV input (dashed curves of Fig. 4(b)). One can
find a similar suppression at low $k_t$ as the two-scale
unintegrated gluon distribution.

    (iii) The drop occurs from factor B, but it is covered by a softer
input sea quark distribution. The dropping position
$k^2_D(x)\rightarrow 0$ in the two-scale unintegrated sea quark
distribution using the MD-DGLAP equation (solid curves of Fig. 4(b))
is in this example, where we take a softer sea quark input
distribution (see Fig. 1 and Eq. (13)).

    We noted that some works [17,18] have used the KMR scheme with the GRV input
to calculate exclusive processes. Unfortunately, these works have
not presented their unintegrated gluon distribution in the form
$f_g(x,k^2_t,\mu^2)/k^2_t\sim k^2_t$ as in Fig. 3(b), therefore, the
dropping fact in the unintegrated gluon distribution has been
hidden.

     The $\mu$-dependence of the two-scale unintegrated gluon distribution is
plotted in Fig. 8. The results show a quicker dropping for
$k_t\rightarrow 0$ with increasing $\mu$.

    The differences between the solid and dashed curves
in Figs. 3 and 4 originate from two sources: one is the direct
contributions from the nonlinear terms in the evolution equation;
the other is due to the change of the input conditions under
shadowing corrections. To illustrate the fist effect, we compare in
Figs. 9 and 10 the solutions of the MD-DGLAP equation with and
without shadowing corrections using the same input of Eqs. (11) and
(13) and $\mu^2_0=0.34~GeV^2$. One can find that the direct
nonlinear corrections in the proton are weaker.

    We know that the nuclear target is an ideal laboratory for the
shadowing research, since the gluon recombination corrections are
enhanced as a result of the correlation of gluons belonging to
different nucleons at the same impact on a nuclear target. We
predict the nuclear parton densities in the MD-DGLAP equation, where
the nonlinear terms are multiplied by $A^{1/3}$ [19]. Figures 11 and
12 give the comparisons of the two-scale gluon and sea quark UPDFs
at $\mu=10 GeV$ in Pb (A=208) with that in the proton, respectively.
In the calculation, the same input, i.e., Eqs. (11) and (13) and
$\mu^2_0=0.34GeV^2$ are used.$\footnote{Strictly speaking in the
case of a heavy nuclei the input distribution at a low starting
scale $\mu_0$ is already modified by the shadowing corrections.
However, we have pointed out that these corrections to the
integrated parton distributions are small [19]. Therefore, here we
have used the ${\it same}$ input for the $Pb$ and for the proton in
order to demonstrate the role of gluon recombination in heavy
nucleus.}$ Comparing them with Figs. 9 and 10, one can find that the
nuclear shadowing effect is important at $x<10^{-4}$. The geometric
scaling implies the following rescaled parton distributions

$$f_a(x,k^2_t,\mu^2)=f_a(x/k^2_t,\mu^2). \eqno(24)$$
We find that the geometric scaling is not evident in our results,
although the $x$ dependence of $f_s(x,k^2_t,\mu^2)$ for the heavy
nucleus in Fig. 12 is roughly flat at the small-$x$ limit. It means
that the nuclear shadowing behavior is far from the saturation
limit.

\newpage
\begin{center}
\section{Comparisons with the other parameterizations of UPDFs}
\end{center}

    Depending on the various QCD evolution dynamics, different
unintegrated gluon distributions are proposed. As an example, we
make some comparisons of our results with the JS (Jung, Salam)
[20,21], KMR (Kimber, Martin, Ryskin) [5], JB (Bl{\"u}mlein) [22]
and GBW (Golec-Bier, W{\"u}sthoff) [23] models.

    The JS gluon is evolved by using the CCFM equation in a Monte Carlo method.
The KMR gluon uses the MRST input distributions and the linear part
of Eqs. (16)-(18) but adding the BFKL-$\ln 1/x$ corrections. The JB
gluon assumes that the integrated and unintegrated gluon
distributions can be connected by using a universal function, which
comes from the expansion of the BFKL anomalous dimension. The
comparisons of our (RZ) result with the JS , JB and KMR models are
illustrated in Fig. 13, where we use a logarithmic scale to indicate
$f_g(x,k^2_t,\mu^2)/k^2_t$. The JS, JB and KMR curves are taken from
Ref. [1], in which the flat part of the KMR curve is an assumption
but not the evolution result, since the MRST input starts from
$\mu^2_0\simeq 1~GeV^2$. We emphasize that the curves in Fig.13
correspond to the different starting scales and different inputs.

        Compared with the JS gluon, the RZ and KMR gluons
have a high-$k_t$ tail at small $x$ in Fig. 13, and it can be
explained by different factorization schemes. The quark distribution
in the RZ and KMR models evolves according to the MD-DGLAP, or DGLAP
equations in the collinear factorization scheme. On the other hand,
the quark evolution in the JS model with the $k_t$ factorization
scheme, is generated by the convolution with off-shell matrix
elements for boson-gluon fusion. This additional evolution broadens
the $k_t$ spectrum of the quarks. Therefore, the JS gluon needs a
narrow $k_t$ spectrum to fit $F_2$ data. Obviously, these
differences arise from different evolution dynamics, and we need
more exclusive data to check their predictions.

        We find that both the JS and JB gluons tend to
rise when $k_t\rightarrow 0$ at small $x$ (although the calculation
of Ref. [22] shows that the JB gluon suddenly becomes negative at
very low $k^2_t\sim 0.01~GeV^2$), whereas the RZ result is
asymptotically dropping toward low $k_t$. As we have emphasized,
this behavior is irrelevant to the saturation effect. In fact,
saturation is thought to arise from the gluon fusion in some
nonlinear evolution equations. The GBW model [23] is one of the
models describing saturation using a color-dipole approach. This
model parameterizes saturation using the single-scale unintegrated
gluon distribution. In Fig. 14 we compare our results (solid curves)
obtained using Eq. (23) with the GBW gluon density (dashed curves).
One can see that the GBW gluon also vanishes for $k_t\rightarrow 0$
due to the nonlinear shadowing effects. On the other hand, the GBW
gluon is strongly suppressed for large $k_t$ values, since the
parton evolution is not treated in the GBW model.

    We noted that a recent work [24] fits low-$Q^2$ dijet data from the
H1 experiment and determines the parameters of the unintegrated
gluon density function by using the linear CCFM CASCADE Monte Carlo
event generator. It is interesting that compared with their old
results, which were determined by fitting $F_2$ data (dashed curves
in Fig. 15), the new parameter (dotted curves in Fig. 15(b))
presents a similar decreasing distribution toward low $k_t$ as our
prediction. Note that the curves in Fig. 15 correspond to the
different scales and different inputs.

        We know that whether a hard valence-like input gluon
distribution or a soft one should be taken in the DGLAP evolution is
still being argued. Our statement gives a way to test different
input models through the observation of the low-$k_t$ behavior of
the unintegrated gluon distribution. Although the application of
perturbative QCD at a very lower-$k_t$ region should be done
carefully, we find that dropping occurs at low but still
perturbative values ($k^2_D>1~GeV^2$ for $x<10^{-3}$, see Fig.6). We
therefore believe that the GRV model provides a possible connection
with the nonperturbative region. For example, a vanished
unintegrated gluon distribution at $k^2_t\rightarrow 0$ is necessary
to obtain a stable $F_L(x,Q^2)$ in $k_t$ factorization [25] and it
seems to favor our results with the GRV input.

    Since particle production at hadronic collisions is sensitive to the
gluon distribution, the various unintegrated gluon distributions are
used to explain the particle multiplicities at the RHIC and predict
the LHC physics. In this aspect, the search for signatures of gluon
saturation effects is a subject of active research. The UPDFs that
include a saturation model are used to explain the particle
multiplicities at the RHIC, and the results are regarded as an
important evidence of saturation effects [17,18,26] In the next
section we give a prediction of our (RZ) UPDFs for the inclusive
gluon distribution.

\newpage
\begin{center}
\section{Prediction of UPDFs for the inclusive gluon distribution}
\end{center}

    First, we consider the cross section for
inclusive gluon production in $proton+proton\rightarrow g$ through
the gluonic mechanism $gg\rightarrow g$ at sufficiently high energy
[9,17], that is

$$\frac{d\sigma}{dyd^2p_t}=\int I(\varphi)d\varphi, \eqno(25)$$

$$I(\varphi)=\frac{4N_c}{N_c^2-1}\frac{1}{p_t^2}
\int\frac{q_tdq_t}{k^2_{1t}k^2_{2t}}\alpha_s(\Omega^2)
f_{1g}(x_1,k^2_{1t},p^2_t)f_{2g}(x_2,k^2_{2t},p^2_t),\eqno(26)$$
where $\Omega^2=\max(k^2_{1t},k^2_{2t},p^2_t)$ ,
$k^2_{1,t}=\frac{1}{4}(p^2_t+q^2_t+2p_tq_t\cos\varphi)$ and
$k^2_{2,t}=\frac{1}{4}(p^2_t+q^2_t-2p_tq_t\cos\varphi)$. The
rapidity $y$ of the produced gluon in the center-of-mass frame of pp
collision is defined by energy-momentum conservation$^{27}$

$$x_{1/2}=\frac{p_t}{\sqrt{s}}\exp(\pm y).\eqno(27)$$

    In the computation, an extrapolation of the unintegrated gluon distribution
at $k^2_t\rightarrow 0$ is needed. This region is beyond the
perturbative QCD framework. Considering the decreasing behavior of
our distribution at $1~GeV^2<k^2_t<k^2_D$ in Fig. 6, we assume

$$f_g(x,k^2_t,p^2_t)/k^2_t=Ak^2_t,~~~at~k^2_t<\mu^2_0, \eqno(28)$$
where $A$ is a parameter connecting the two parts of $f_g$ at
$k^2_t>\mu^2_0$ and $k^2_t<\mu^2_0$.

    In Fig. 16, we show the intrinsic angular correlation function
$I(\varphi)$  for our model (solid curve) of unintegrated gluon
distribution at RHIC energy $W=200~GeV$. In this calculation $y=0$
and $p_t = 1~GeV$ were taken. A similar correlation function but
using the GBW saturation model is also presented (dotted curve). We
also give the result of Ref. [17] (dashed curve), which also uses
the KMR scheme with the GRV model, but the flattened distribution of
$f_g(x, k^2_t, p_t^2)/k^2_t= const.$ at $k^2_t<0.5~GeV^2$ is
assumed. One can find that a quite different (oscillatory or flat)
pattern is obtained for dropping or not-dropping behaviors of the
unintegrated gluon distribution at low $k_t$. The rapidity
distribution is presented in Fig. 17, where the integration over
$p_t^2>0.34~GeV^2$ was performed.

    In experiments, a good identification of particles is not always
achieved, which makes it hard to determine the rapidity of a
particle. The practice then is to measure pseudorapidity. The
rapidity and pseudorapidity distributions of partons for massless
particles are identical. The situation changes when massive
particles are produced in the final state via fragmentation and in
this case the rapidity $y$ can be obtained from the pseudorapidity
$\eta$

$$y=\frac{1}{2}\ln \left[\frac{\sqrt{\frac{m^2_{eff}+p^2_t}{p^2_t}+\sinh^2 \eta}+\sinh \eta}
{\sqrt{\frac{m^2_{eff}+p^2_t}{p^2_t}+\sinh^2\eta}-\sinh \eta}
\right], \eqno(29)$$ where $m_{eff}$ is the mass of the typical
produced hadron.$^{26}$ Assuming that pions in pp collision are
produced via $\rho$-resonance, we take $m_{eff}=770~MeV$.

    To avoid the complicated hadronization dynamics, similar to Ref. [28], we
use local parton-hadron duality, that is, the rapidity distribution
of particles is identical to the rapidity distribution of gluons.
Thus, the pseudorapidity density of the produced charged particles
in proton-proton collisions is given by

$$\frac{1}{\sigma_{in}}\frac{d \sigma(\eta, p_t)}{d\eta d^2p_t}=CJ(\eta;p_t;m_{eff})\left.\frac{d \sigma(y,p_t)}{dy
d^2p_t}\right|_{y=\eta}, \eqno(30)$$ where $J(\eta;p_t;m_{eff})$ is
jacobian

$$J(\eta;p_t;m_{eff})=\frac{\cosh\eta}{\sqrt{\frac{m^2_{eff}+p^2_t}{p^2_t}+\sinh^2\eta}}, \eqno(31)$$
and the gluon to charged hadron ratio via gluon fragmentation model
is accounted for by the normalization constant $C$, which is fixed
by the pseudorapidity spectrum of $pp\rightarrow \pi^{\pm} X$ at
$W=200~MeV$ and $\eta=0$. Corrections to the kinematics due to the
hadron mass are also considered by replacing $p^2_t\rightarrow
p^2_t+m^2_{eff}$ in the evaluation of $x_{1/2}$. Figure 18 gives our
prediction, where  $C=0.0153~GeV^2$. The experimental data of the
UA5 collaboration are taken from Ref. [29]. One can find a sinking
platform, which was presented in various saturation models [17,28],
but our results are irrelevant to the saturation mechanism.

    Our theoretical spectrum in Fig. 18 is too steep to agree with
experimental data in the fragmentation region. We noted that the
pseudorapidity distribution of charged pions using a saturated
unintegrated gluon distribution in the $gg\rightarrow g$ mechanism
was calculated by Kharzeev and Livin (the KL model) [26] and their
results described well the RHIC data. However, contrary to the claim
of Ref. [26], a similar calculation was reproduced by Szczurek, who
used same KL model in Ref. [17] and his theoretical pseudorapidity
distributions are still significantly lower than experimental data
in the fragmentation region. This difference originates from the
different assumptions used about the unintegrated gluon distribution
at the larger $x$ region. In fact, the unintegrated gluon
distribution in the KL model is valid with certainty only for $x <
0.1$. To extrapolate the gluon distribution to $x>0.1$, two
different multiple factors $(1-x)^4$ and $(1-x)^{5-7}$ in the gluon
distribution are assumed in Refs. [26] and [17], respectively. This
uncertainty is now removed, since the UPDFs of our model are well
defined in the whole $x$ and $k_t$ space.

    It was well known that the recombination of the original fast
quarks of two incident hadrons with a slow antiquark cannot be
neglected in the fragmentation region [30] Our results in Fig. 16
are compatible with Szczurek's work [17] and they suggest the
involvement of other hadronization mechanism in the fragmentation
regions. The specificity of this effect will be discussed elsewhere.

    Now we use the above results to predict the
inclusive pion distribution in the central region ($\vert
\eta\vert<2$) at LHC energies. The result for the pseudorapidity
density of charged pion in central p-p collisions at $W=5.5~TeV$ is
shown in Fig. 19. The dashed curves remind us that the quark
recombination effects in the fragmentation region are neglected.
Comparing the plots with those in Fig. 18, we find that the width of
the rapidity distribution increases with increasing energy $W$. The
reason is that the unintegrated gluon distribution is enhanced
toward smaller $x$ without saturation behavior in the $x$
distribution in the RZ UPDFs. Thus, either $x_1$ or $x_2$ decreases
fast with increasing $W$ at a higher fixed value of $y$, and it
leads to broadening of the rapidity distribution with increasing
$W$. It is different from our model in that the $x$ dependence of
the unintegrated gluon in the KL model is saturated to a constant at
small $x$ and it almost keeps the width of the rapidity
distribution, as shown in Ref. [28].

   In summary, the unintegrated parton distributions in the proton and heavy nucleus
are predicted by using the MD-DGLAP equation incorporating the
shadowing corrections, which include exact energy-momentum
conservation in each splitting and fusion vertex. We find that the
nuclear shadowing effects are obvious although they are far from the
saturation limit. On the other hand, we point out that the
suppression of the unintegrated gluon distribution at lower $k_t^2$
may arise from the valence-like input rather than the saturation
effects.

\noindent {\bf Acknowledgments}: This work was supported by the
National Natural Science Foundations of China, No. 10875044.

\newpage

\noindent {\bf Figure Captions}

\noindent Fig. 1  Two valence-like input distributions for sea
quark: our modified input Eq. (13) (solid curve) and GRV98LO input
Eq.(12)(dashed curve). Note that different starting scales $\mu^2_0$
are used.

\noindent Fig. 2  The fits of the computed $F_2(x,Q^2)$ in proton by
the MD-DGLAP equation (solid curves) comparing with H1 and ZEUS
data. The dashed curves are the DGLAP equation results with GRV98LO
input.

\noindent Fig. 3 The unintegrated, two-scale dependent gluon
distribution in proton at $\mu=10~GeV$ (a) as a function of $x$ for
different values of $k^2_t$; (b) as a function of $k_t$ for
different values of $x$. The solid (or dashed) curves are the
solutions of the MD-DGLAP (or DGLAP) equations, where two different
inputs are used.  $k^2_D$ is the dropping scale, where the
distribution begins suppression towards lower $k^2_t$.

\noindent Fig. 4 The two-scale unintegrated sea quark distribution
in proton at $\mu=10~GeV$ (a) as a function of $x$ for different
values of $k^2_t$; (b) as a function of $k_t$ for different values
of $x$. The solid (or dashed) curves are the solutions of the
MD-DGLAP (or DGLAP) equations, where two different inputs are used.

\noindent Fig. 5 As same as Fig. 4 but for the valence quark
distributions.

\noindent Fig. 6 The $x$-dependence of the dropping scale $k^2_D(x)$
for the two-scale unintegrated gluon distribution (solid curve). The
dashed curve is the saturation scale according to
$Q^2_s(x)=1~GeV^2(10^{-4}/x)^{0.277}$.

\noindent Fig. 7 Similar to Fig. 3 but for the single-scale
unintegrated gluon distribution using Eq. (23).

\noindent Fig. 8 The evolution scale $\mu^2$-dependence of the
two-scale unintegrated gluon distribution at $x=10^{-6}$ for
different values of $k^2_t$.

\noindent Fig. 9 Similar to Fig. 3 but using the same input Eqs.
(11), (13) and $\mu^2_0=0.34~GeV^2$.

\noindent Fig. 10 Similar to Fig. 4 but using the same input Eqs.
(11), (13) and $\mu^2_0=0.34~GeV^2$.

\noindent Fig. 11 Comparison of the two-scale unintegrated gluon
distributions in Pb(A=208) (solid curves) with proton (dashed
curves) at $\mu=10~GeV$, (a) for $x$-dependence and (b) for
$k_t$-dependence.

\noindent Fig. 12 As same as Fig. 11 but for the two-scale
unintegrated sea quark distribution.

\noindent Fig. 13 Comparison of our predicted (RZ)-gluon
distribution (solid curves) with KMR-gluon (dashed curves), JS-gluon
(dotted curves) and JB-gluon (broken-dotted curves), (a) for
$x$-dependence and (b) for $k_t$-dependence. The dashed and dotted
curves are taken from Ref. [1]. Note that the flatten part of the
KMR curves is an assumption in Ref. [1] but not the evolution
result.

\noindent Fig. 14 The single-scale unintegrated gluon distribution
using Eq. (23) as a function of $k_t^2$ for different values of $x$
(solid curves) and the comparison with the saturation GBW model [23]
(dashed curves). Note that the GBW gluon lacks a larger $k_t$ tail
since the parton evolution is not treated in this model.

\noindent Fig. 15 Comparison of our two-scale unintegrated gluon
distribution (solid curves) with two results of the CCFM equation in
Ref. [24], which are obtained by fitting $F_2$ data (dashed curves)
and low $Q^2$ di-jet data (dotted curves), respectively.

\noindent Fig. 16 The intrinsic azimuthal correlations for different
unintegrated gluon distributions: our (RZ) mode (solid curve), GBW
model (dotted curve) and Szczurek results [17] (dashed curve).

\noindent Fig. 17 Inclusive gluon rapidity distribution ($p_t^2 >
0.34~GeV^2)$ at $W= 200~GeV$ for our unintegrated gluon
distribution.

\noindent Fig. 18 Pseudo-rapidity density of charged pion produced
in p-p collisions at $W=~200 GeV$. Data are taken from Ref. [29].

\noindent Fig. 19 Predicted pseudo-rapidity density of charged pion
produced in p-p collisions at $W= 5.5~TeV$. The dashed curves remind
us that the quark recombination effects in the fragmentation region
are neglected.

\newpage

 \hbox{

\centerline{\epsfig{file=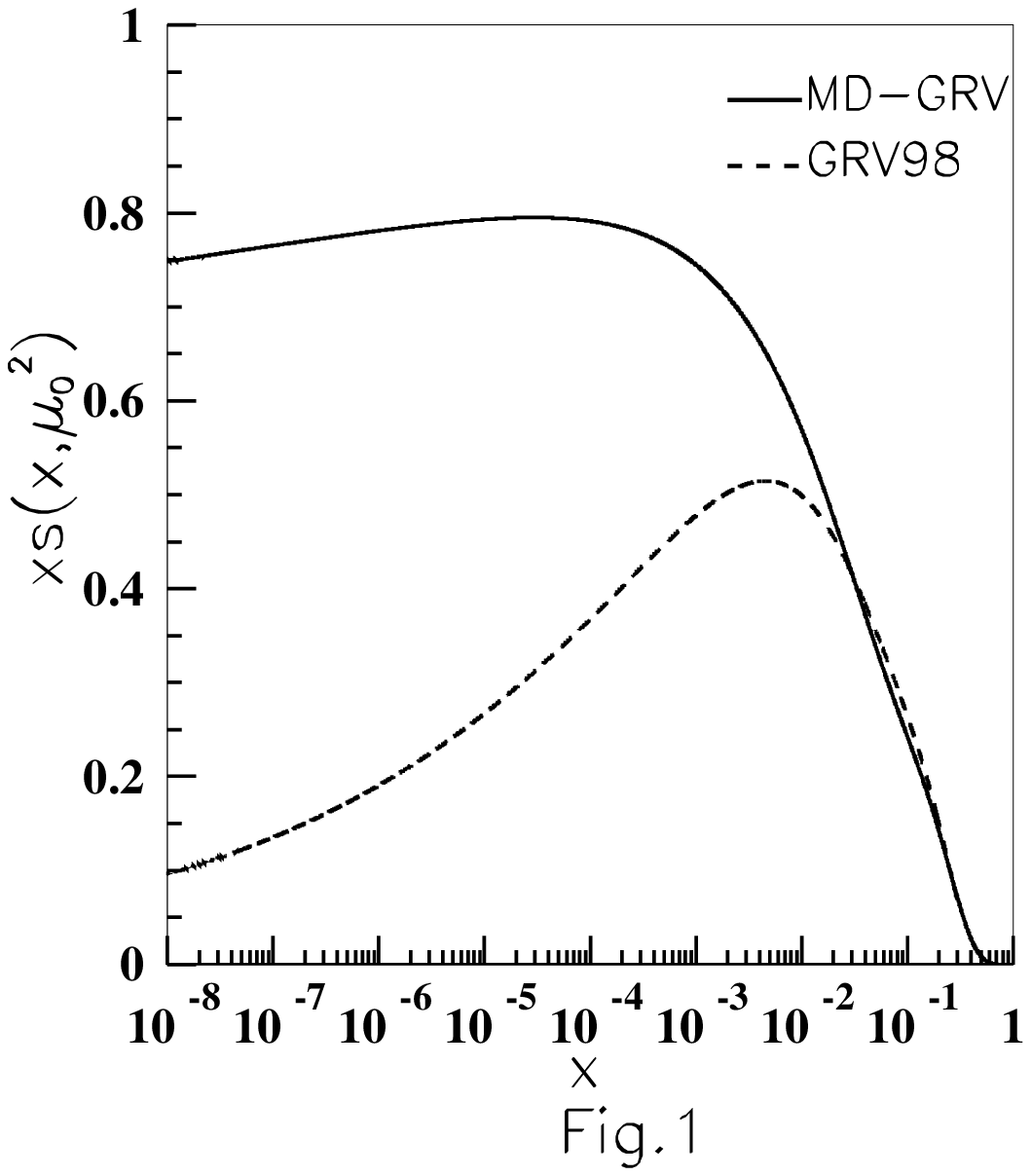,width=15.0cm,clip=}}}

 \hbox{

\centerline{\epsfig{file=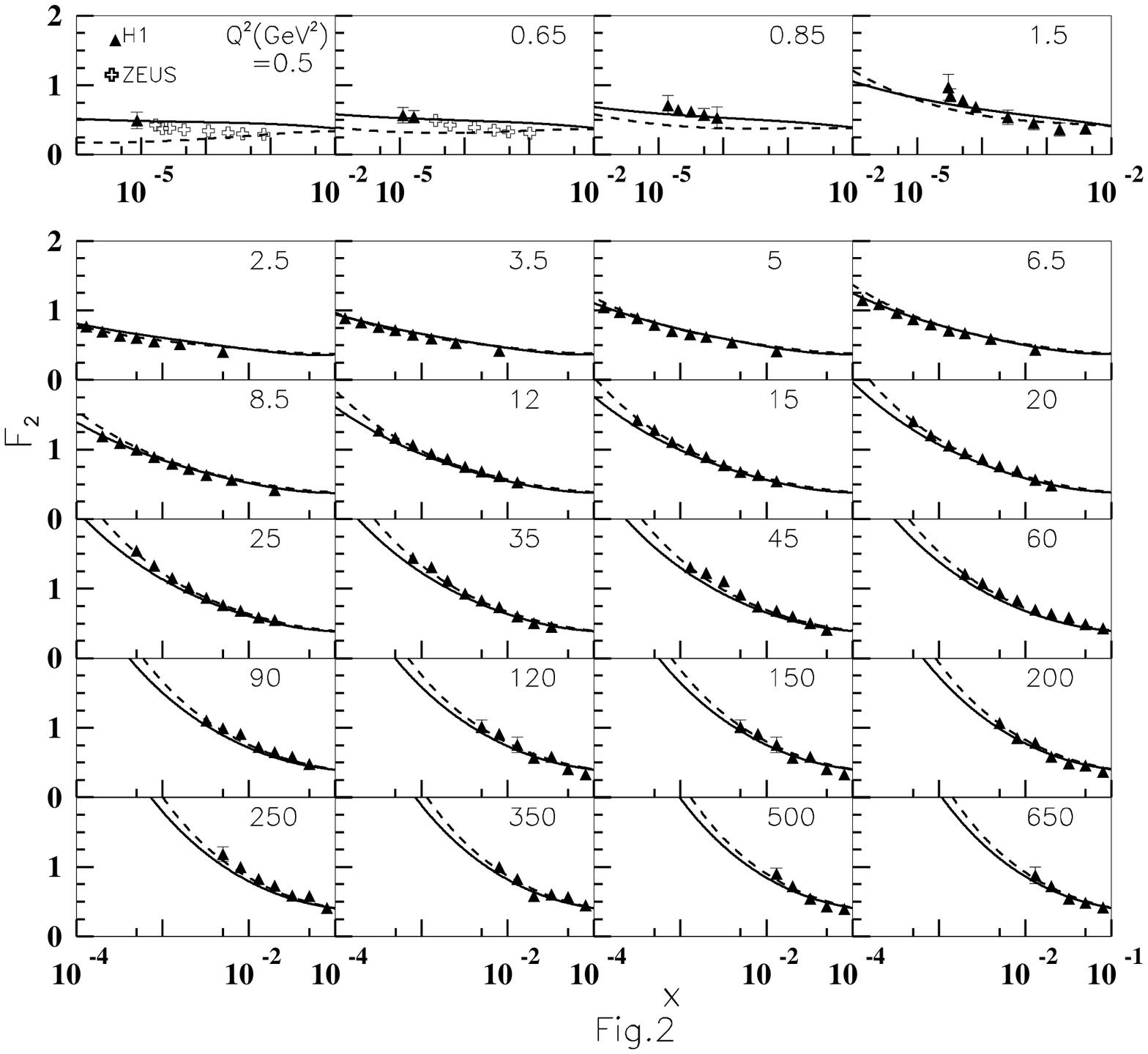,width=15.0cm,clip=}}}

 \hbox{

\centerline{\epsfig{file=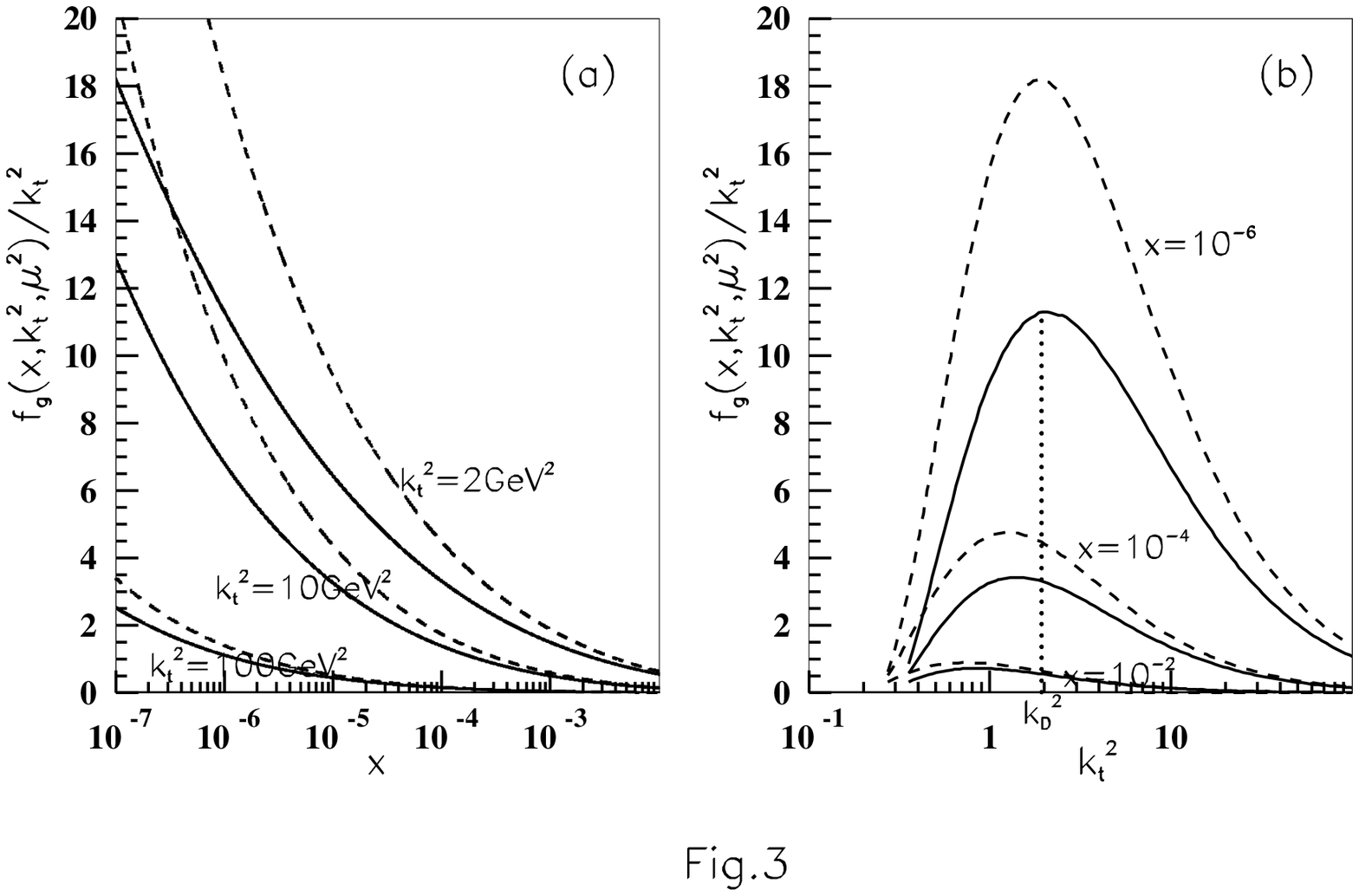,width=15.0cm,clip=}}}

 \hbox{

\centerline{\epsfig{file=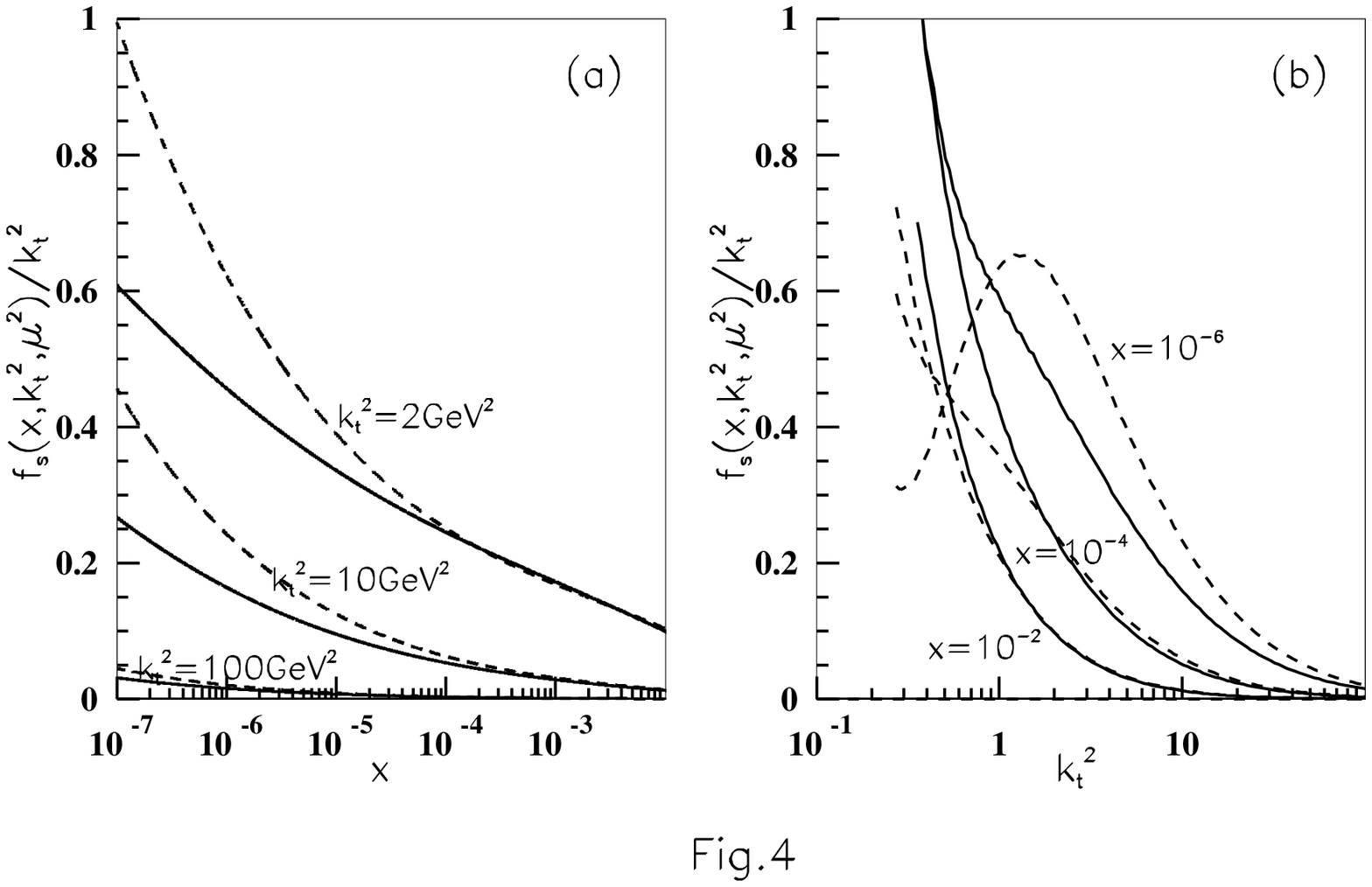,width=15.0cm,clip=}}}

 \hbox{

\centerline{\epsfig{file=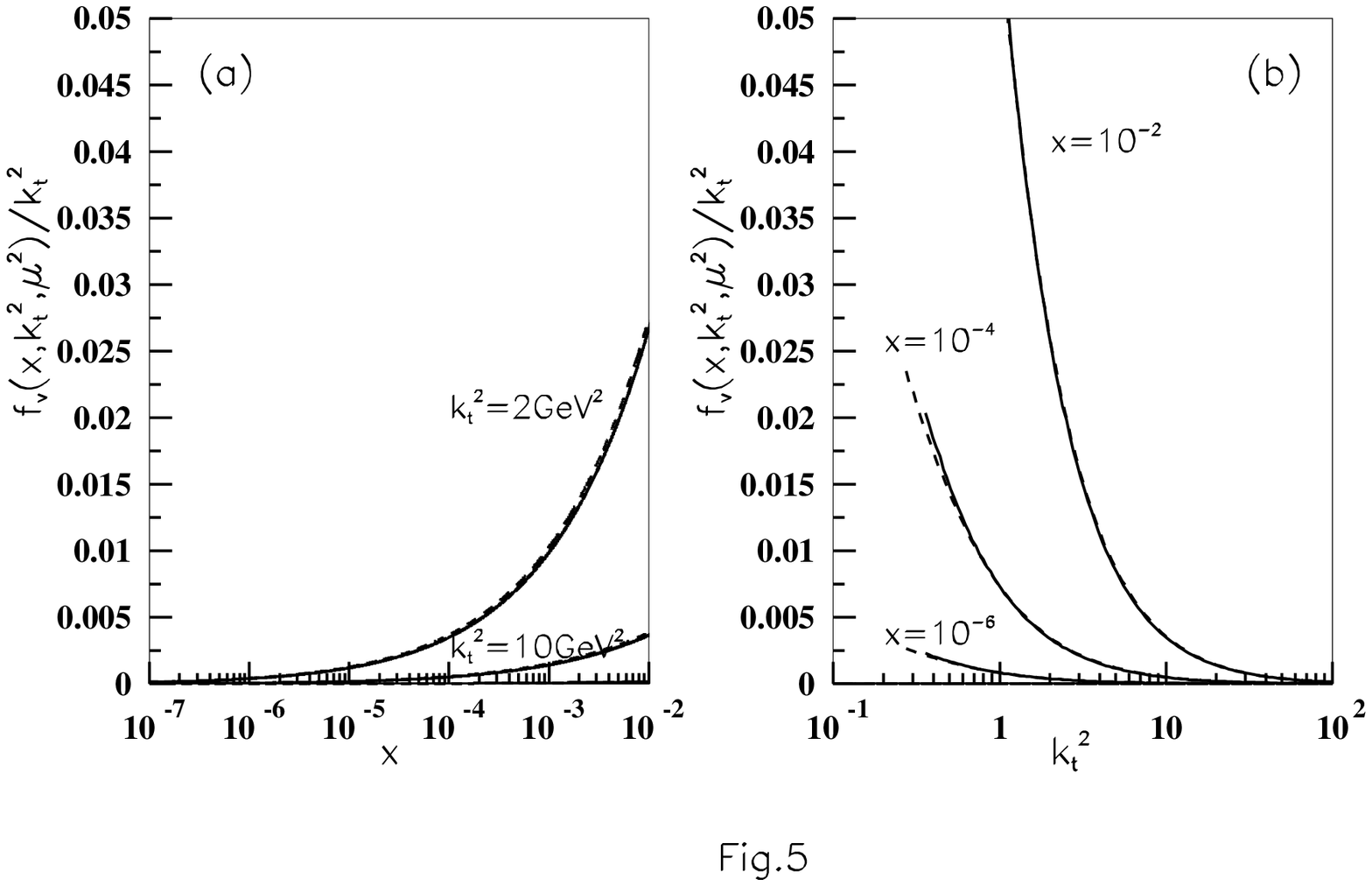,width=15.0cm,clip=}}}

 \hbox{

\centerline{\epsfig{file=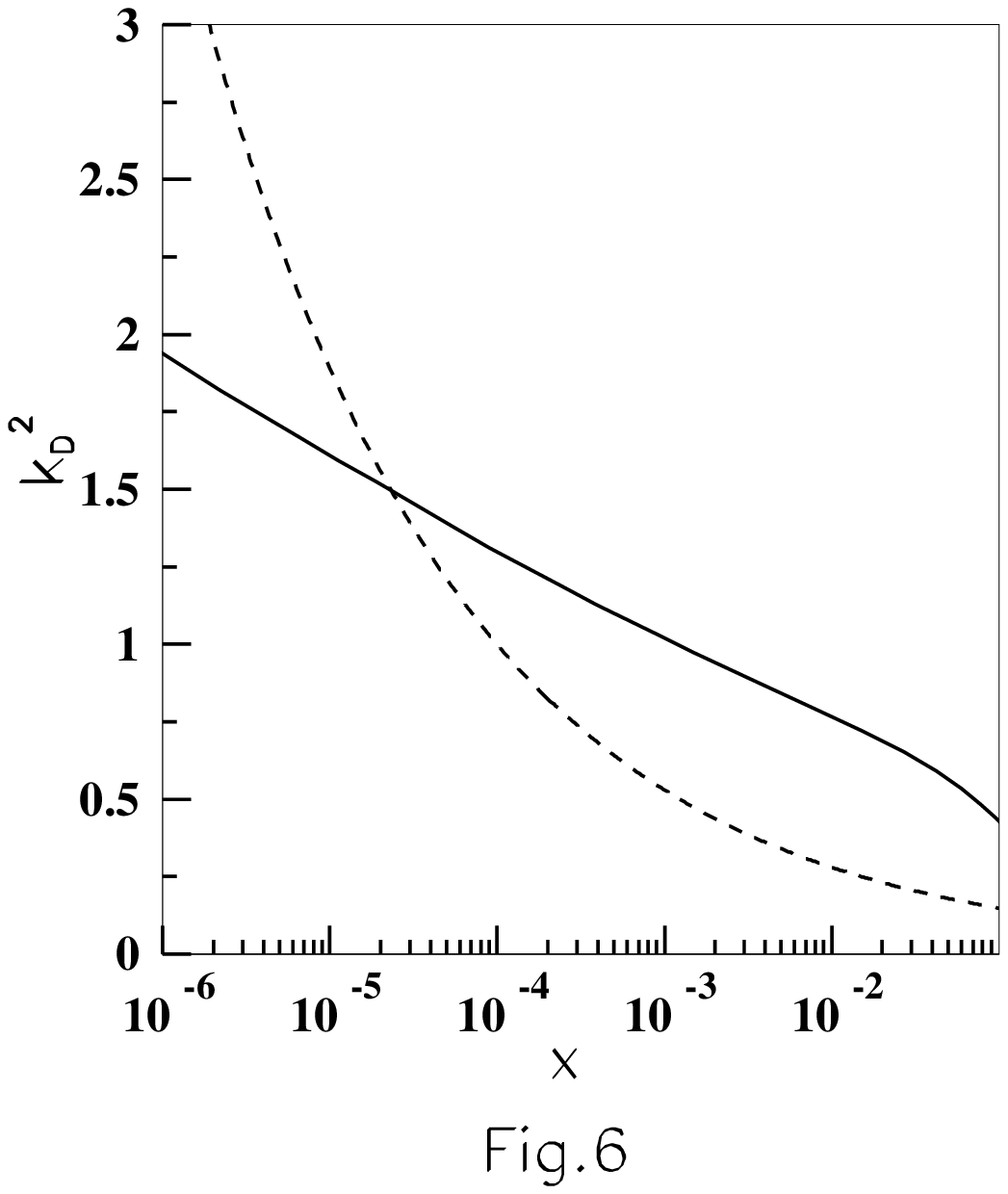,width=15.0cm,clip=}}}

 \hbox{

\centerline{\epsfig{file=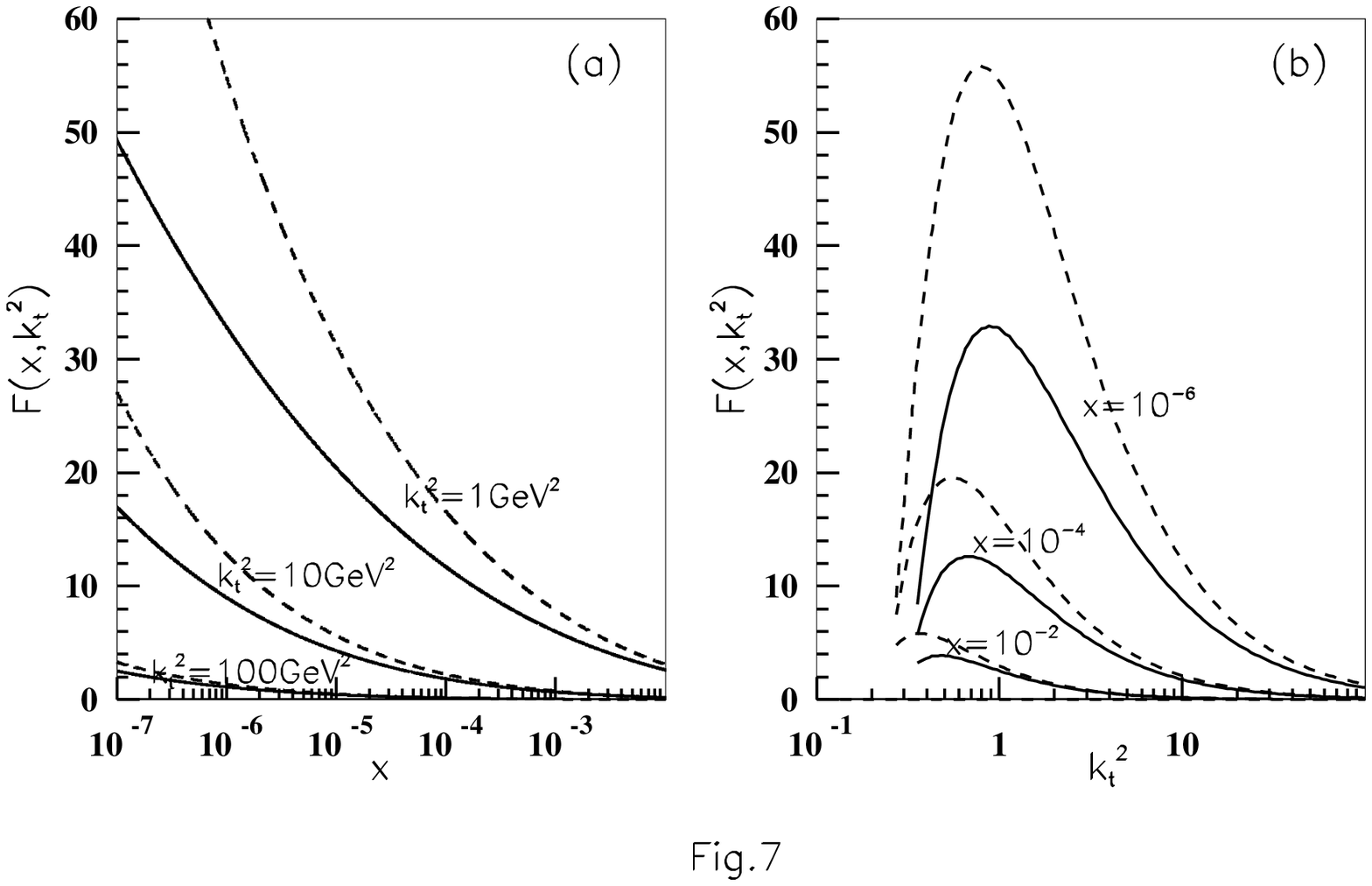,width=15.0cm,clip=}}}

 \hbox{

\centerline{\epsfig{file=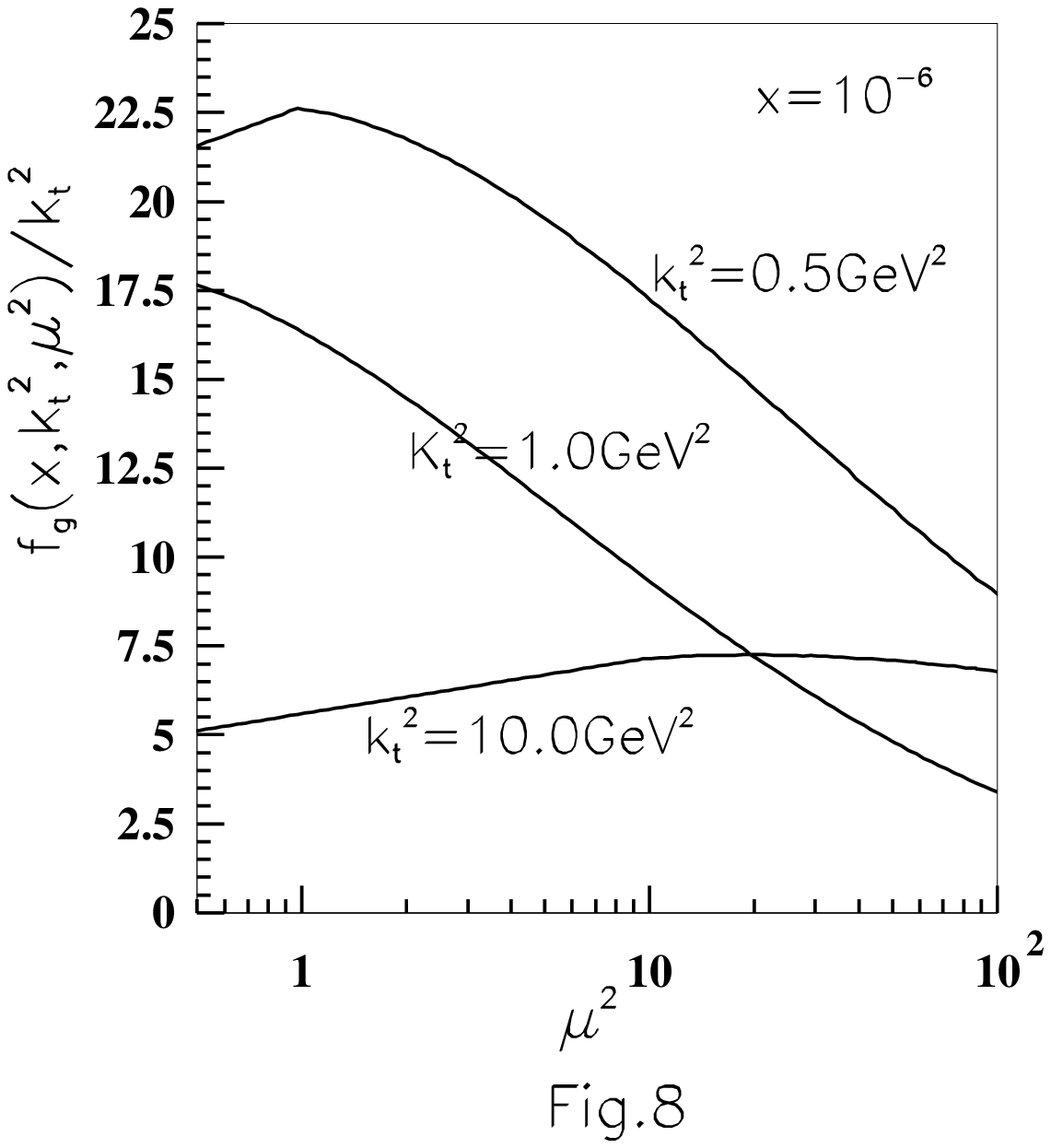,width=15.0cm,clip=}}}
 \hbox{

\centerline{\epsfig{file=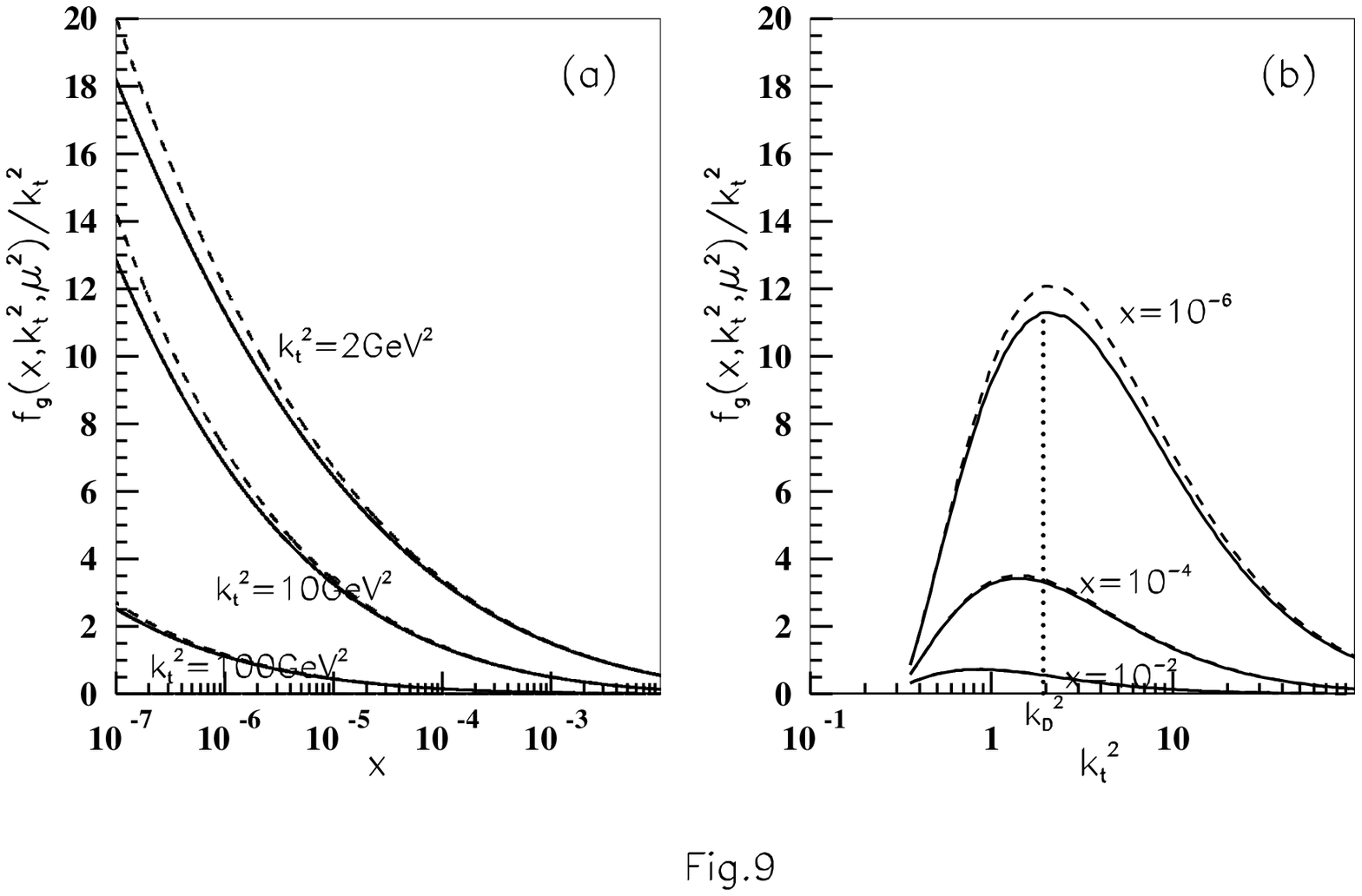,width=15.0cm,clip=}}}
 \hbox{

\centerline{\epsfig{file=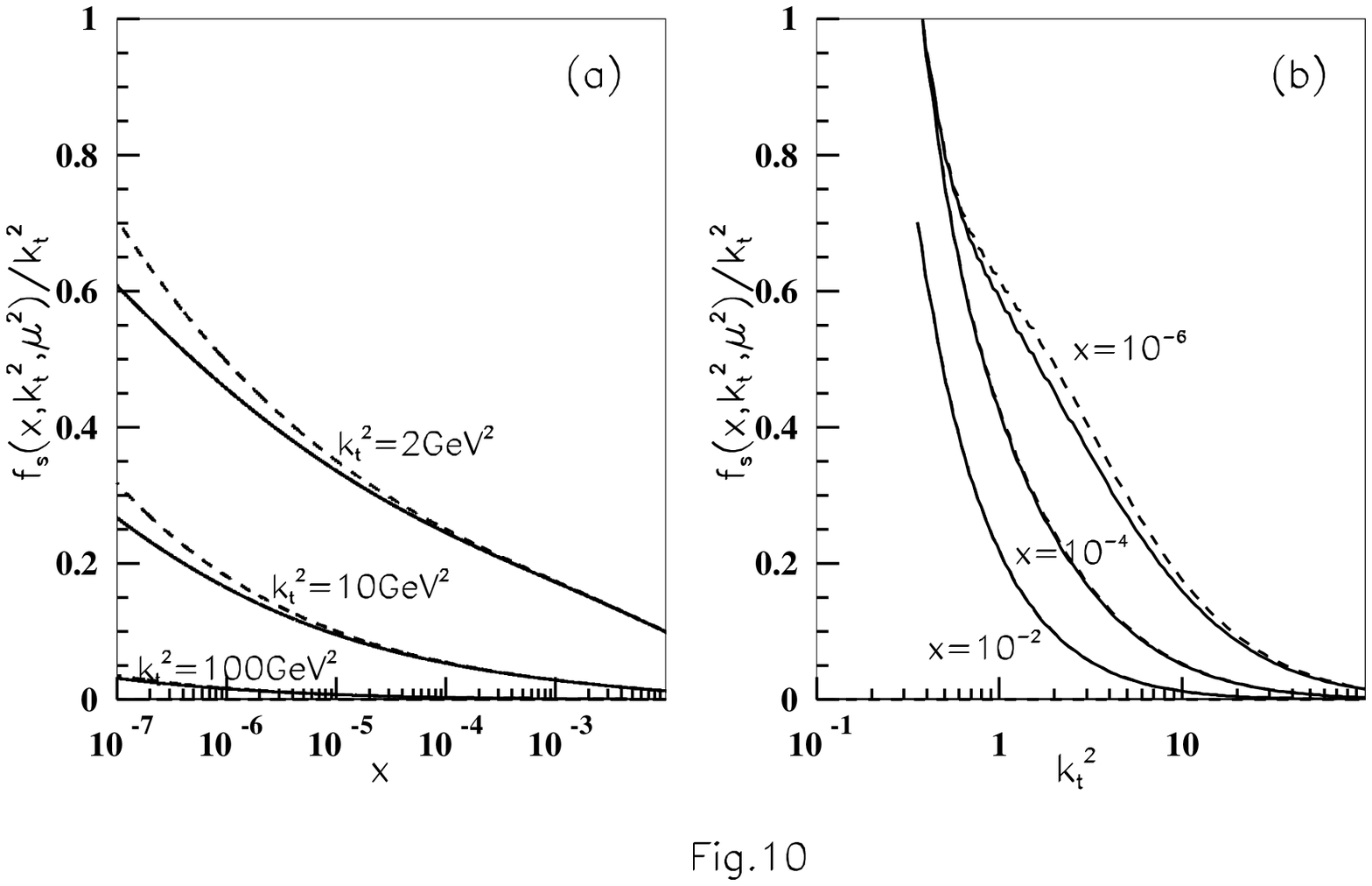,width=15.0cm,clip=}}}
 \hbox{

\centerline{\epsfig{file=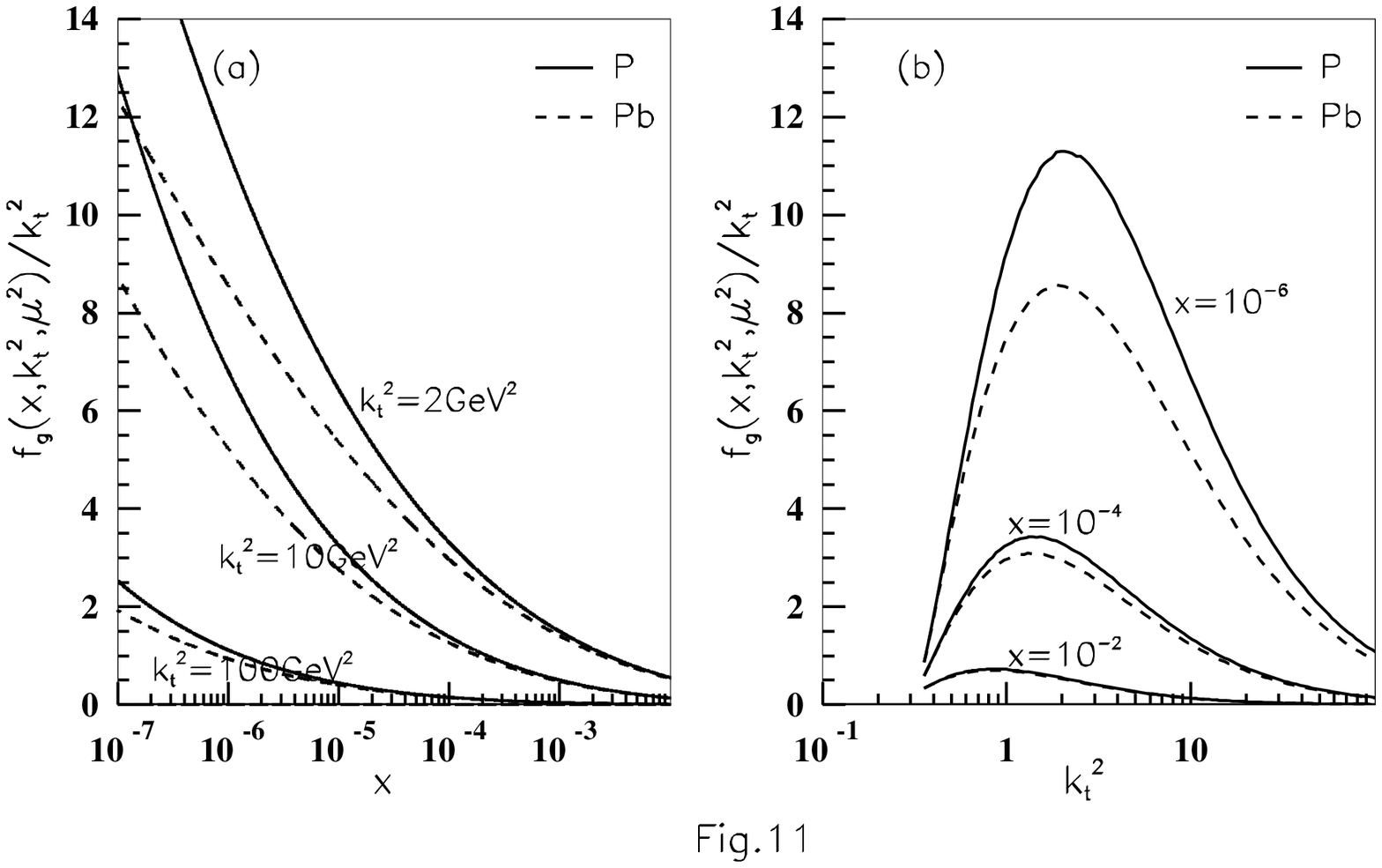,width=15.0cm,clip=}}}
 \hbox{

\centerline{\epsfig{file=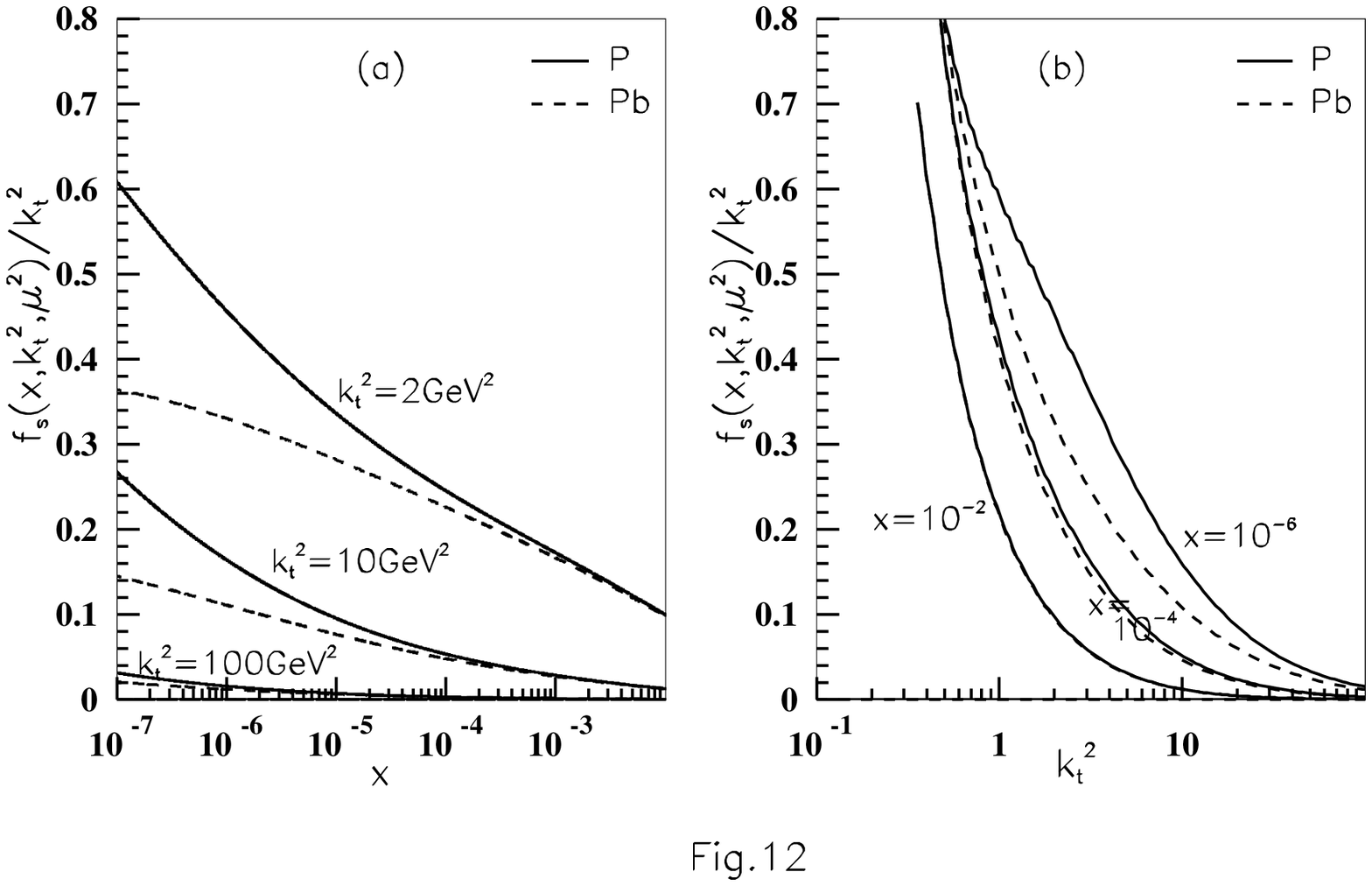,width=15.0cm,clip=}}}
 \hbox{

\centerline{\epsfig{file=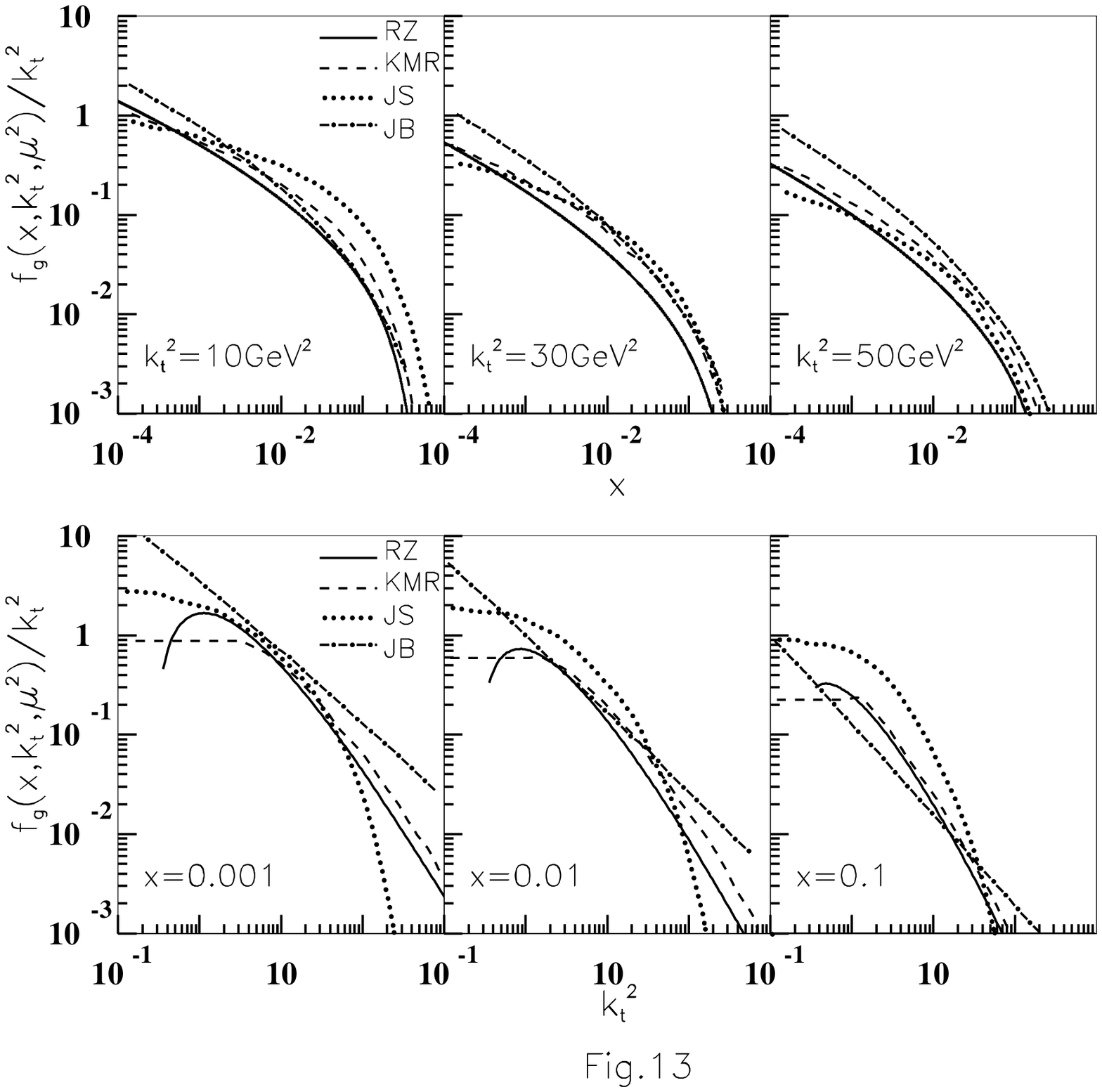,width=15.0cm,clip=}}}
 \hbox{

\centerline{\epsfig{file=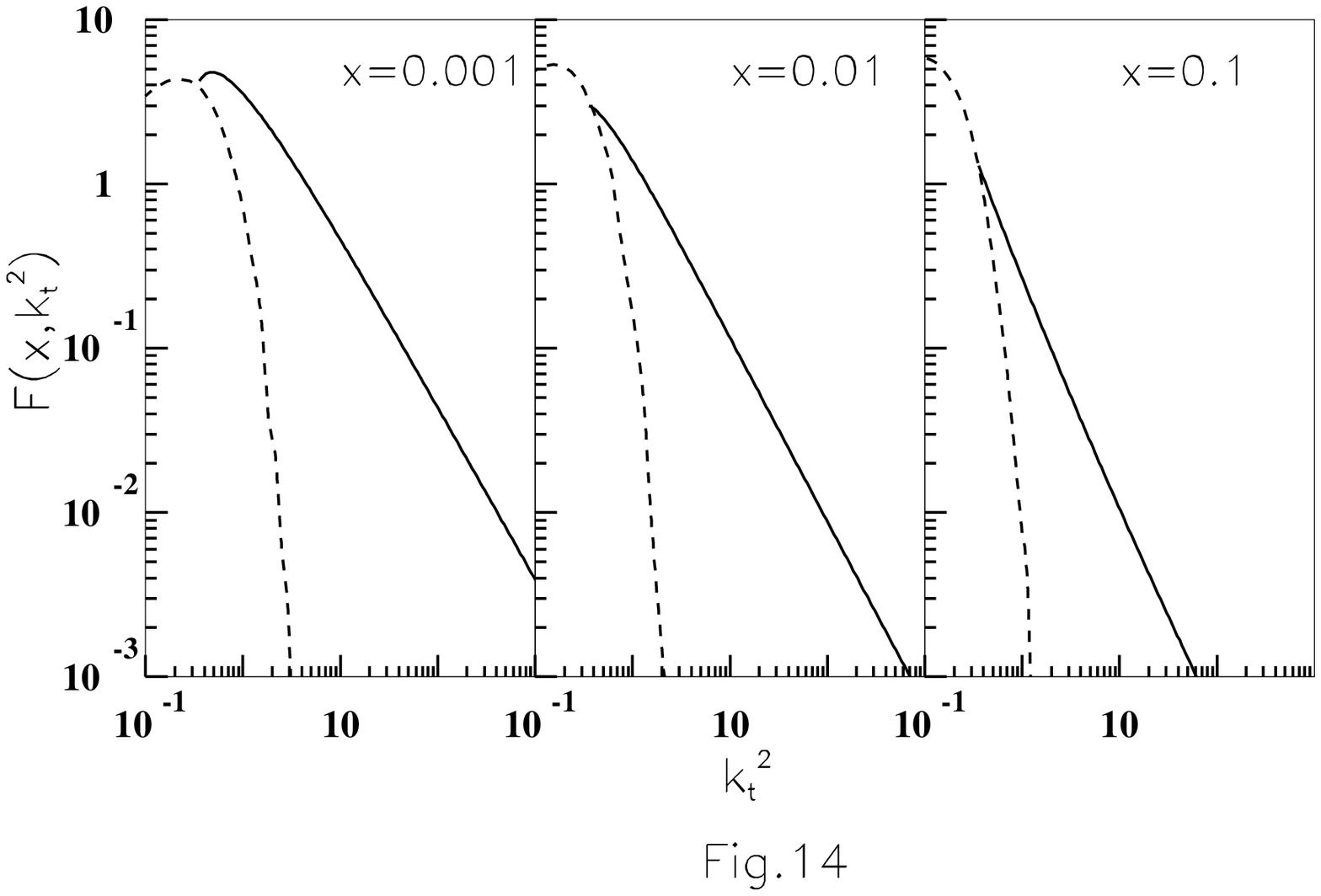,width=15.0cm,clip=}}}
 \hbox{

\centerline{\epsfig{file=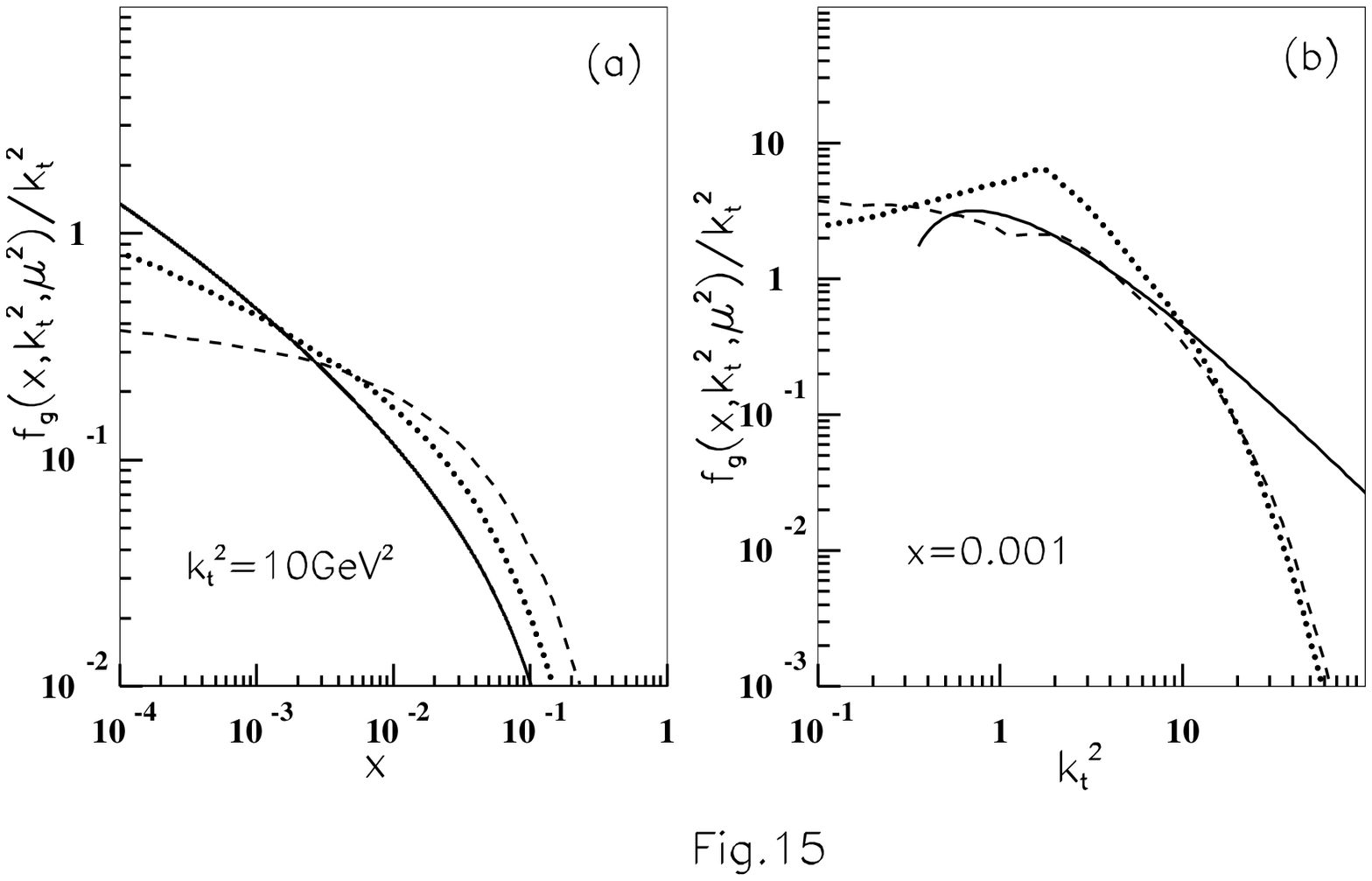,width=15.0cm,clip=}}}
 \hbox{

\centerline{\epsfig{file=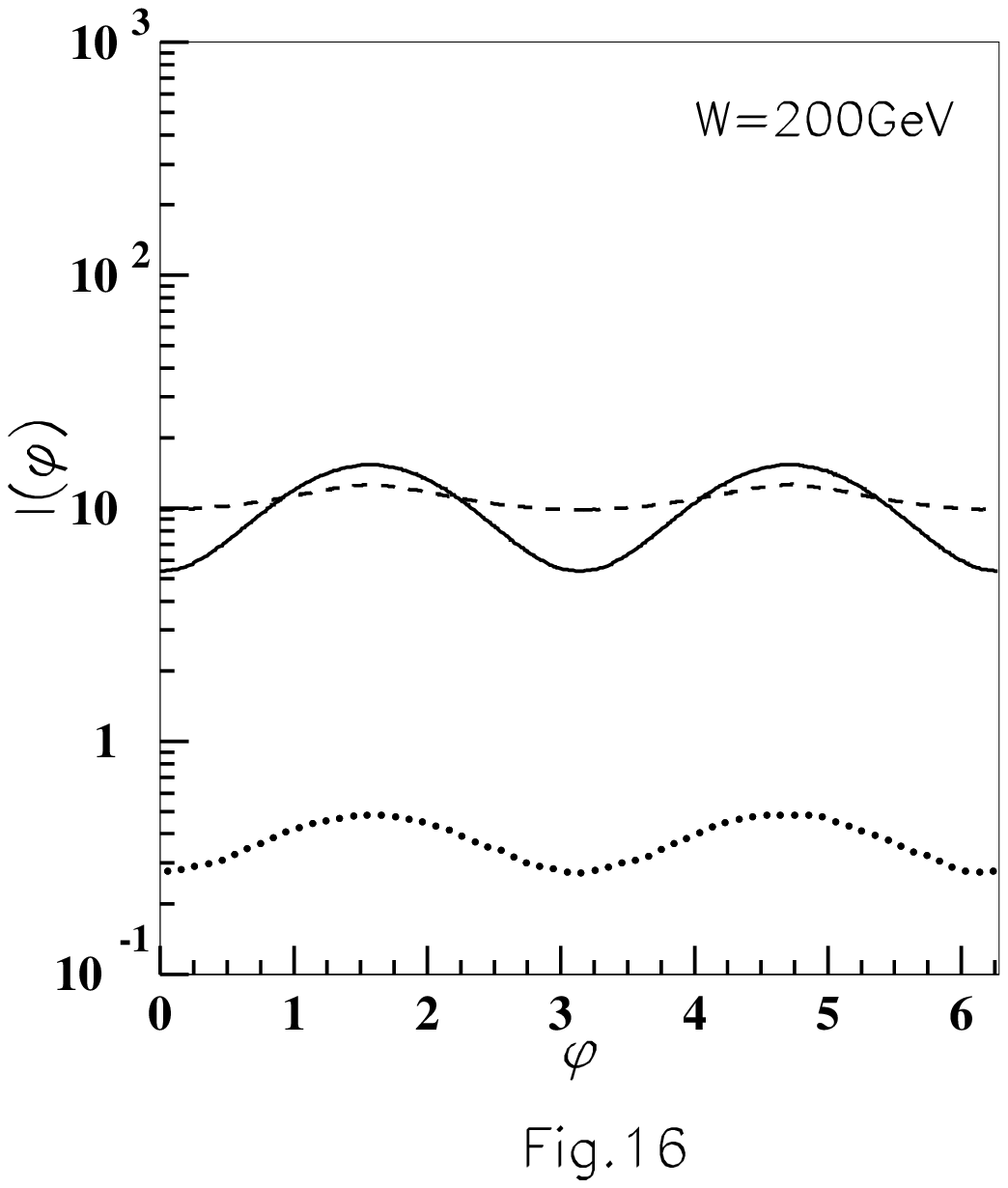,width=15.0cm,clip=}}}
 \hbox{

\centerline{\epsfig{file=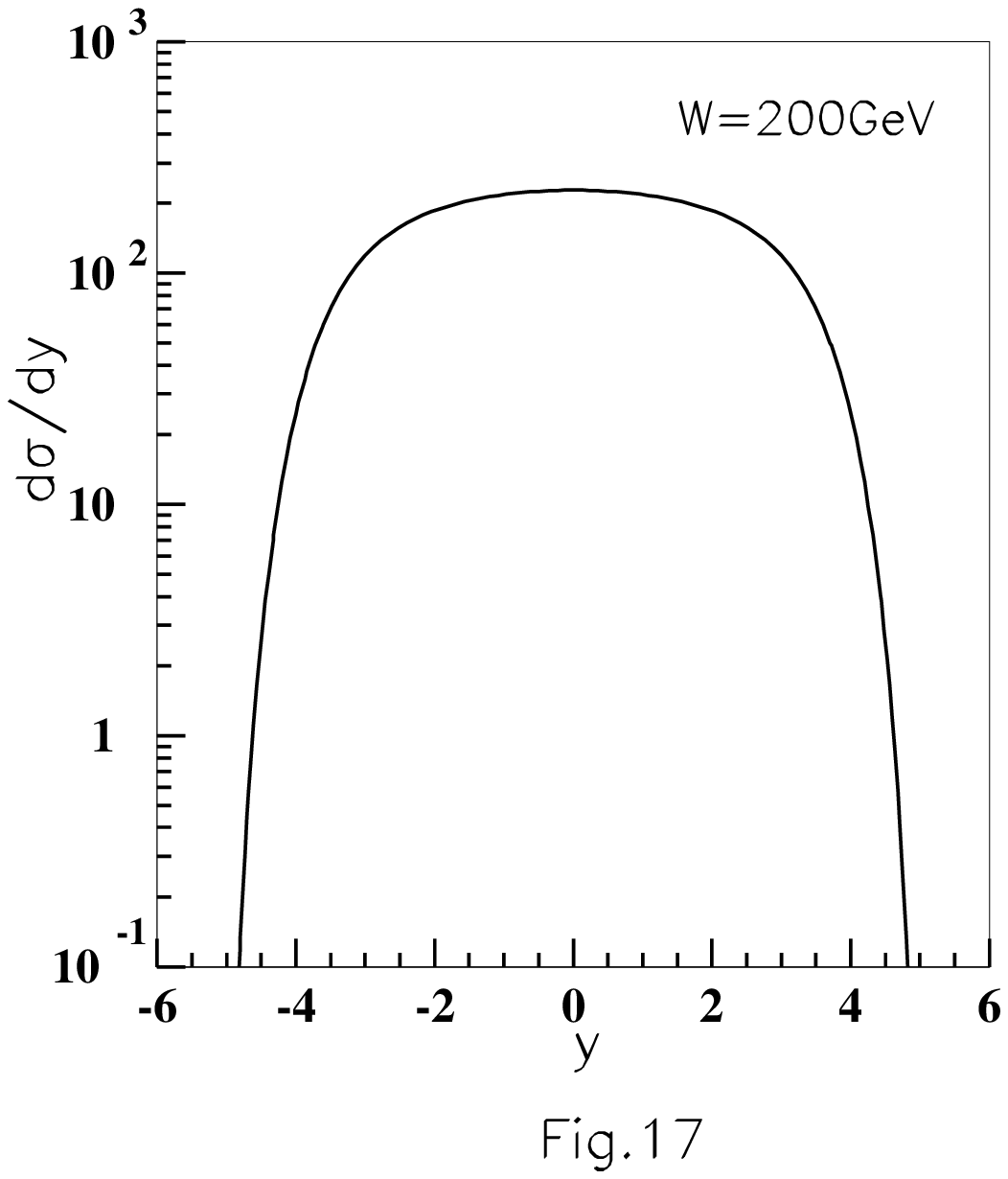,width=15.0cm,clip=}}}
 \hbox{

\centerline{\epsfig{file=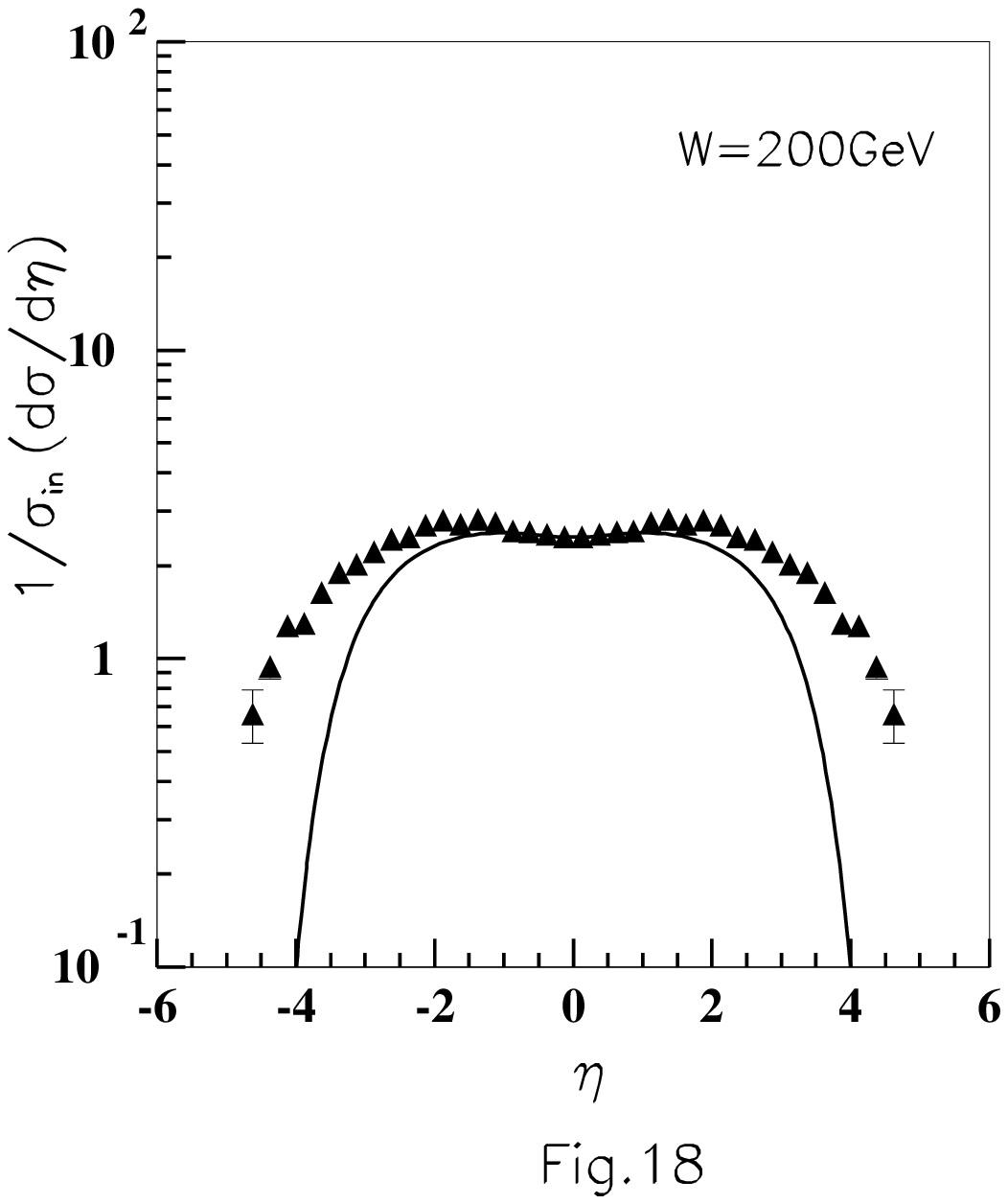,width=15.0cm,clip=}}}
 \hbox{

\centerline{\epsfig{file=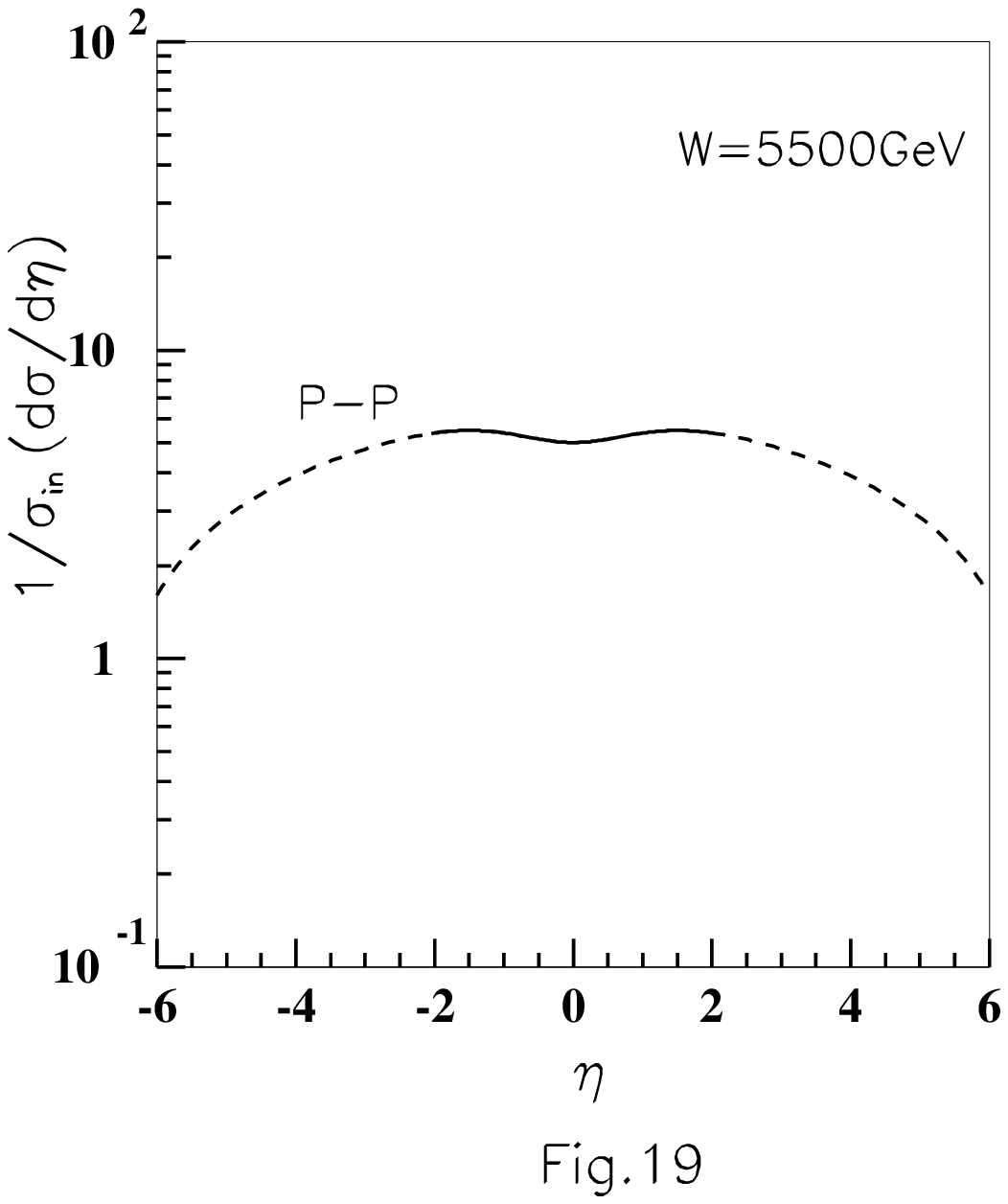,width=15.0cm,clip=}}}


\end{document}